\documentclass[aps,prl,preprint,nopacs,superscriptaddress]{revtex4}
\usepackage{graphicx}
\usepackage{verbatim}
\usepackage{relsize}
\usepackage{mathrsfs}
\usepackage{color}
\usepackage{ulem}
\usepackage{amsmath,amsfonts,amssymb}
\usepackage{graphicx}
\usepackage{braket}
\definecolor{mygray}{gray}{0.6}
\usepackage{bbding}
\usepackage{pifont}
\usepackage{wasysym}
\usepackage{amssymb}
\usepackage{multirow}

\def \beq {\begin{equation}}
\def \eeq {\end{equation}}
\begin{document}

\title{\normalsize A Chiral Superlattice Route to Spin-Split Topological Antiferromagnetism
}

\author{\scriptsize Thao Dinh$^*$}\affiliation{\scriptsize Department of Chemistry and Chemical Biology, Harvard University, MA 02138, USA}\affiliation{\scriptsize Department of Physics, Harvard University, Cambridge, MA 02138, USA}
\author{\scriptsize Mengke Liu\footnote{These authors contributed equally to this work.}$^{\dag}$}\affiliation{\scriptsize Department of Physics, University of Texas at Dallas, Richardson, Texas 75080, USA}\affiliation{\scriptsize Department of Physics, Harvard University, Cambridge, MA 02138, USA}
\author{\scriptsize Jian-Xiang Qiu$^{\dag}$}\affiliation{\scriptsize Department of Chemistry and Chemical Biology, Harvard University, MA 02138, USA}

\author{\scriptsize Sougata Mardanya}\affiliation{\scriptsize Department of Physics and Astronomy, Howard University, Washington, DC 20059, USA}
\author{\scriptsize Suk Hyun Sung}\affiliation{\scriptsize The Rowland Institute at Harvard, Harvard University, Cambridge, MA 02138, USA}
\affiliation{\scriptsize Michigan Center for Materials Characterization, University of Michigan, Ann Arbor, MI 48105}

\author{\scriptsize Xuan Hoang Le}\affiliation{\scriptsize Department of Chemistry and Chemical Biology, Harvard University, MA 02138, USA}\affiliation{\scriptsize Department of Physics, Harvard University, Cambridge, MA 02138, USA}

\author{\scriptsize Christopher Broyles}\affiliation{\scriptsize Department of Physics, Washington University in St. Louis, St. Louis, MO 63130, USA}

\author{\scriptsize Chengfeng Zhu}\affiliation{\scriptsize Department of Chemistry and Chemical Biology, Harvard University, MA 02138, USA}

\author{\scriptsize Qiaozhi Xu}\affiliation{\scriptsize Department of Physics, Washington University in St. Louis, St. Louis, MO 63130, USA}
\author{\scriptsize Haotian Chen}\affiliation{\scriptsize Department of Physics, Washington University in St. Louis, St. Louis, MO 63130, USA}
\author{\scriptsize Zack Rehfuss}\affiliation{\scriptsize Department of Physics, Washington University in St. Louis, St. Louis, MO 63130, USA}

\author{\scriptsize Yu-Fei Liu}\affiliation{\scriptsize Department of Chemistry and Chemical Biology, Harvard University, MA 02138, USA}\affiliation{\scriptsize Department of Physics, Harvard University, Cambridge, MA 02138, USA}

\author{\scriptsize Xiaoyu Zeng}\affiliation{\scriptsize Department of Chemistry and Chemical Biology, Harvard University, MA 02138, USA}

\author{\scriptsize Houchen Li}\affiliation{\scriptsize Department of Chemistry and Chemical Biology, Harvard University, MA 02138, USA}

\author{\scriptsize Peng Guo}\affiliation{\scriptsize Department of Chemistry and Chemical Biology, Harvard University, MA 02138, USA}
\author{\scriptsize Jinchen Liu}\affiliation{\scriptsize Department of Chemistry and Chemical Biology, Harvard University, MA 02138, USA}
\author{\scriptsize Tianye Huang}\affiliation{\scriptsize Department of Chemistry and Chemical Biology, Harvard University, MA 02138, USA}

\author{\scriptsize Jingtian Shi}\affiliation{\scriptsize Materials Science Division, Argonne National Laboratory, Lemont, IL 60439, USA}
\author{\scriptsize Michael Smith}\affiliation{\scriptsize Materials Science Division, Argonne National Laboratory, Lemont, IL 60439, USA}
\author{\scriptsize Joanna M. Blawat}\affiliation{\scriptsize National High Magnetic Field Laboratory, Los Alamos National Laboratory, Los Alamos, New Mexico 87545, USA}
\author{\scriptsize John Singleton}\affiliation{\scriptsize National High Magnetic Field Laboratory, Los Alamos National Laboratory, Los Alamos, New Mexico 87545, USA}
\author{\scriptsize Ross McDonald}\affiliation{\scriptsize National High Magnetic Field Laboratory, Los Alamos National Laboratory, Los Alamos, New Mexico 87545, USA}

\author{\scriptsize Iv\'an E. Arvizo}\affiliation{\scriptsize Department of Chemistry and Chemical Biology, Harvard University, MA 02138, USA}

\author{\scriptsize Dongtao Cui}\affiliation{\scriptsize Department of Chemistry and Chemical Biology, Harvard University, MA 02138, USA}

\author{\scriptsize Shao-Liang Zheng}\affiliation{\scriptsize Department of Chemistry and Chemical Biology, Harvard University, MA 02138, USA}

\author{\scriptsize Kenji Watanabe}\affiliation{\scriptsize Research Center for Electronic and Optical Materials, National Institute for Materials Science, 1-1 Namiki, Tsukuba 305-0044, Japan}

\author{\scriptsize Takashi Taniguchi}\affiliation{\scriptsize Research Center for Materials Nanoarchitectonics, National Institute for Materials Science, 1-1 Namiki, Tsukuba 305-0044, Japan}

\author{\scriptsize Vineet Kumar Sharma}\affiliation{\scriptsize Department of Physics and Astronomy, Howard University, Washington, DC 20059, USA}
\author{\scriptsize Sudip Ghorai}\affiliation{\scriptsize S. N. Bose National Centre for Basic Sciences, Kolkata, India}
\author{\scriptsize Barun Ghosh}\affiliation{\scriptsize S. N. Bose National Centre for Basic Sciences, Kolkata, India}

\author{\scriptsize Hsin Lin} \affiliation{\scriptsize Institute of Physics, Academia Sinica, Taipei 11529, Taiwan}

\author{\scriptsize Ting Yong Lim}\affiliation{\scriptsize Department of Physics, National Cheng Kung University, Tainan 701, Taiwan}

\author{\scriptsize Tay-Rong Chang}\affiliation{\scriptsize Department of Physics, National Cheng Kung University, Tainan 701, Taiwan}

\author{\scriptsize Arun Bansil}\affiliation{\scriptsize Department of Physics, Northeastern University, Boston, MA 02115, USA}\affiliation{\scriptsize Quantum Materials and Sensing Institute, Northeastern University, Burlington, MA 01803, USA
}

\author{\scriptsize Ivar Martin}\affiliation{\scriptsize Materials Science Division, Argonne National Laboratory, Lemont, IL 60439, USA}

\author{\scriptsize Ashvin Vishwanath}\affiliation{\scriptsize Department of Physics, Harvard University, Cambridge, MA 02138, USA}

\author{\scriptsize Weiwei Xie}\affiliation{\scriptsize Department of Chemistry, Michigan State University, MI 48824, USA}
\author{\scriptsize Claudia Felser}\affiliation{\scriptsize Max Planck Institute for Chemical Physics of Solids, Dresden, Germany.}
\author{\scriptsize Jairo Sinova}\affiliation{\scriptsize Institut f\"ur Physik, Johannes Gutenberg University, 55099 Mainz, Germany}
\affiliation{\scriptsize Department of Physics, Texas A\&M University, College Station, Texas 77843-4242, USA}

\author{\scriptsize Anyuan Gao}\affiliation{\scriptsize Department of Chemistry and Chemical Biology, Harvard University, MA 02138, USA}
\author{\scriptsize  Mikhail D. Lukin}\affiliation{\scriptsize Department of Physics, Harvard University, Cambridge, MA 02138, USA}
\author{\scriptsize Jennifer E. Hoffman}\affiliation{\scriptsize Department of Physics, Harvard University, Cambridge, MA 02138, USA}

\author{\scriptsize Sugata Chowdhury}\affiliation{\scriptsize Department of Physics and Astronomy, Howard University, Washington, DC 20059, USA}

\author{\scriptsize Qiong Ma}\affiliation{\scriptsize Department of Physics, Boston College, Chestnut Hill, MA, USA}
\author{\scriptsize Kai Sun}\affiliation{\scriptsize Department of Physics, University of Michigan Ann Arbor, Ann Arbor, MI, 48109, USA}
\author{\scriptsize Ismail El Baggari}\affiliation{\scriptsize The Rowland Institute at Harvard, Harvard University, Cambridge, MA 02138, USA}
\author{\scriptsize Sheng Ran}\affiliation{\scriptsize Department of Physics, Washington University in St. Louis, St. Louis, MO 63130, USA}
\author{\scriptsize Talieh S. Ghiasi}\affiliation{\scriptsize Department of Physics, Harvard University, Cambridge, MA 02138, USA}\affiliation{\scriptsize Kavli Institute of Nanoscience, Delft University of Technology, Delft, The Netherlands}

\author{\scriptsize Hongkun Park}\affiliation{\scriptsize Department of Chemistry and Chemical Biology, Harvard University, MA 02138, USA}\affiliation{\scriptsize Department of Physics, Harvard University, Cambridge, MA 02138, USA}

\author{\scriptsize Philip Kim$^{\dag}$}\affiliation{\scriptsize Department of Physics, Harvard University, Cambridge, MA 02138, USA}

\author{\scriptsize Su-Yang Xu\footnote{Corresponding authors: suyangxu@fas.harvard.edu, philipkim@g.harvard.edu, mliu@utdallas.edu, jqiu@g.harvard.edu}}\affiliation{\scriptsize Department of Chemistry and Chemical Biology, Harvard University, MA 02138, USA}

\maketitle


\textbf{The interplay between chirality and magnetism is a profound topic, spanning particle physics, condensed matter, chemistry and origin of life \cite{pasteur1862researches, barron2000chirality, juraschek2025chiral}. In this paper, we study this interplay in the new antiferromagnet (AFM) UOTe, and we report a chiral superlattice route to spin-split topological phenomena from collinear antiferromagnetism. Pristine UOTe is a collinear AFM without Berry curvature or spin-split bands. However, we identify the spontaneous formation of a chiral superlattice in UOTe, which can be viewed as frozen chiral phonons at finite wave vector. When electrons move through the chiral superlattice, their orbital Bloch wavefunction and quantum geometry are modulated by the strong chiral superlattice potential, generating large Berry curvature in the chiral superlattice, which we detect by the nonlinear Hall effect. At 150 K, the chiral superlattice induced Berry curvature couples to the collinear AFM order, generating a giant anomalous Hall angle $\sim0.14$ abruptly near $T_\textrm{N}$ that is among the largest in bulk magnets. Moreover, our spin Hanle precession measurement shows that the chiral superlattice also generates spin-polarized current from the collinear AFM, a long-standing goal in spintronics. We shed light on the supermodulation formation mechanism based on chemical ion size and physical interlayer-intralayer energy competitions, using which we propose a design principle to discover similar bond-mismatch superlattices, providing a chiral superlattice pathway through real space engineering of quantum geometry \cite{dong2018geometrodynamics,yao2025dynamical}.}

\vspace{0.5cm}
\textbf{Introduction}

The concept of superlattice has recently emerged at the forefront of quantum condensed matter physics, highlighted by the 2D vdW moir\'e heterostructures. In addition to the artificially-constructed superlattices, some bulk materials can, intriguingly, develop intrinsic long range lattice modulation, forming natural superlattice. Such naturally-occurring superlattices offer potential advantages in structural uniformity and scalability \cite{gao1988incommensurate}. Very recently have researchers begun to investigate this direction \cite{devarakonda2020clean,qian2022chiral,devarakonda2024evidence}. Despite the recent interest, the formation mechanism of natural superlattice is not well understood - most material discoveries are based on trials and errors. 

In this paper, we report the discovery of a natural chiral superlattice in a collinear AFM UOTe. Collinear AFMs are long thought as promising platform for spintronics, with generating large AHE and spin-polarized current as central goals \cite{jungwirth2016antiferromagnetic, vsmejkal2018topological}, motivating the recent explosion of altermagnetism \cite{jungwirth2026symmetry, chen2014anomalous, smejkal_beyond_2022,mazin_editorial_2022,cheong2025altermagnetism,song2025altermagnets, nakatsuji2015large,  nayak2016large}. From the symmetry point of view, one needs to break the $\mathcal{PT}$ and $t\mathcal{T}$ symmetries. However, experimentally, although researchers have extensively explored many methods, the effect has been weak. We show that the natural chiral superlattice serves as a new and effective way to functionalize the collinear AFM for topological magnetic phenomena distinct from altermagnetism. Unlike altermagnets where spin splitting arises directly from crystal and magnetic symmetry in the collinear AFM \cite{jungwirth2026symmetry, chen2014anomalous, smejkal_beyond_2022,mazin_editorial_2022,cheong2025altermagnetism,song2025altermagnets, nakatsuji2015large,  nayak2016large}, pristine UOTe is a collinear AFM with $\mathcal{PT}$ and $t\mathcal{T}$ symmetries. However, a spontaneous chiral superlattice breaks the relevant symmetries and moreover generates real space quantum geometry, thereby producing altermagnetic-like topological responses. Moreover, we uncover the formation mechanism for the chiral superlattice. Distinct from the moire superlattice, the chiral superlattice in UOTe is a bond-mismatch superlattice, in which the superlattice strongly modifies the local covalent U-Te bond. This led to a strong modulation whose energy scale is roughly one order of magnitude larger than typical moir\'e systems, enabling the real space quantum geometry. We propose a design principle combining chemistry, physics and materials to discover similar bond-mismatch chiral superlattice in other materials. Seminal studies of chiral charge-density wave in the Kagome CsV$_3$Sb$_5$ \cite{yu2021concurrence,guo2022switchable,guo2024distinct,elmers2025chirality} has made important breakthroughs in the context of exotic time-reversal symmetry breaking states such as the loop current order with connection to unconventional superconductivity. In this work, we establish a chiral superlattice route - both mechanistic and materials-based - for realizing spin-split topological responses in collinear AHE, providing both significant results and new microscopic mechanisms in the contexts of topological magnets, quantum geometry and natural superlattice.

\vspace{0.5cm}
\textbf{Basic characterizations}

\begin{figure*}[t]
\includegraphics[width=15cm]{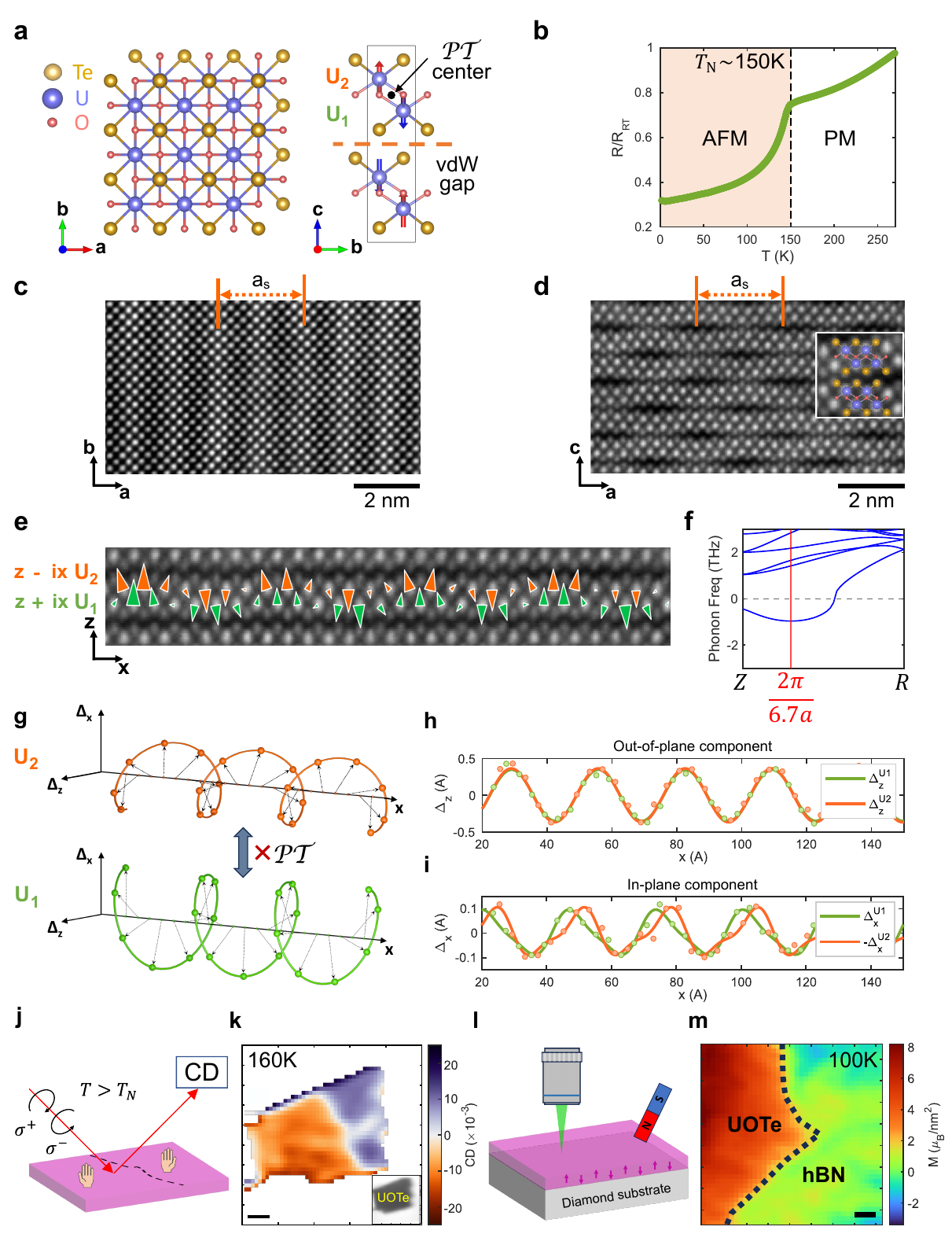}
\vspace{-0cm}
\caption{\textbf{Discovery of natural superlattice with 2D planar chirality in the collinear AFM UOTe.} \textbf{a,} Crystal structure and collinear AFM of UOTe. \textbf{b,} Resistance of UOTe flake ($13.3$ nm) shows metallic temperature dependence. \textbf{c,d,} Plan-view and cross-sectional view of STEM of UOTe. A spontaneous superlattice with in-plane period $a_s \approx 6.7a$ is observed. \textbf{e,} Cross-sectional STEM image of one UOTe vdW layer. The arrows represent the displacement of U1 and U2. Along the supermodulation axis $\hat{x}$, the displacement of U1 rotates within the $(x,z)$ plane from $-x$ to $+z$, then to $+x$, and finally to $-z$ within one period $a_s$. The displacement of U2 also rotates within the $(x,z)$ plane but with the opposite chirality. }
\label{Fig1}
\end{figure*}

UOTe is a new vdW heavy fermion based magnet \cite{broyles_uote_2025}. Structurally, UOTe has a layered, square crystal structure with the space group P4/nmm $\#129$, shown in Fig.~\ref{Fig1}\textbf{a}. Magnetically, UOTe shows a collinear AFM order (Fig.~\ref{Fig1}\textbf{a}) with $T_{\textrm{N}}\simeq150$ K, reported by neutron scattering in Ref. \cite{broyles_uote_2025}. Within the same vdW layer, the two U ions, labeled U1 and U2, are AFM-coupled with their spins along the $+z$ and $-z$ directions. The magnetic unit cell contains two vdW layers, with the spin configuration of $(+z, -z) (-z, +z)$. This collinear AFM has $\mathcal{PT}$ and $t\mathcal{T}$ symmetries ($\mathcal{P}$, $\mathcal{T}$ and $t$ are space-inversion, time-reversal and translation symmetries). Electronically, UOTe has a semimetallic ground state. Figure~\ref{Fig1}\textbf{b} shows the temperature-dependent resistance. The carrier density is low, $2\times10^{19}$ cm$^{-3}$. 

\vspace{0.5cm}
\textbf{Discovery of spontaneous chiral superlattice}

We have performed transmission electron microscopy (TEM) measurements at room temperature. Beyond the atomic structure described above, our TEM data show the emergence of a long-range wavelike superlattice (Figs.~\ref{Fig1}\textbf{c,d}). The superlattice is found to be present up to 700 $^{\circ}$C. The in-plane period of the superlattice is $a_s \approx 2.65~\text{nm}\approx 6.7 a$, therefore incommensurate with the atomic lattice. In reciprocal space, this corresponds to a wavevector $\vec{q}=(\frac{1}{6.7}\frac{2\pi}{a}, 0, \frac{1}{2}\frac{2\pi}{c}$). $q_z= \frac{1}{2}\frac{2\pi}{c}$. Moreover, the superlattice in UOTe is unidirectional, which breaks the four-fold rotational symmetry of the atomic lattice. The two orthogonal superlattice domains, $q_x$ and $q_y$, are observed by low-energy electron microscopy (LEEM) measurements (Supplementary Information (SI).I.2.), with a typical domain size $\sim100$ $\mu$m.

We now visualize the atomic displacement pattern.  Along the supermodulation axis $\hat{x}$, the displacement of U1 - green arrows in Fig.~\ref{Fig1}\textbf{e} - rotates within the $(x,z)$ plane, from $-x$ to $+z$, then to $+x$, and finally to $-z$. Similarly, the displacement of U2 also rotates within the $(x,z)$ plane but with the opposite chirality. Such supermodulations of U1 and U2 can be viewed as the ``freezing'' of chiral phonons \cite{juraschek2025chiral} with opposite handedness: $z+ix$ for U1 and $z-ix$ for U2 (DFT phonon calculations in Fig.~\ref{Fig1}\textbf{f} and Extended Data Fig.~\ref{Extended_DFT_phonon}).  Similarly, the Te1 and Te2 also show the opposite chiral winding structures.

We now show the existence of a net chirality in the superlattice. Our measured $\Delta_{\textrm{U}1 ,x}$ and $\Delta_{\textrm{U}2 ,x}$, have different shapes, as shown in Fig.~\ref{Fig1}\textbf{i}. The different shapes show that the opposite chiral displacements of U1 and U2 do not cancel. Such a net chirality breaks $\mathcal{PT}$ symmetry.  In addition, $t\mathcal{T}$ symmetry is also broken because the in-plane supermodulation of adjacent layers have different lineshapes (Extended Data Fig.~\ref{Extended_TEM_300K_100K}\textbf{b}). It is interesting to note that the lineshapes of different layers in Extended Data Fig.~\ref{Extended_TEM_300K_100K}\textbf{b} suggest (pseudo-)periodicity, which is also reflected in the fitted $\phi_{q}$ versus layer in Extended Data Fig.~\ref{Extended_TEM_300K_100K}\textbf{c}. On the other hand, the diffraction data (SI.Fig.S2) does not show corresponding satellite peak in the $q_z$ direction, excluding long-range ordering. The possible short-range order requires further investigations, but $t\mathcal{T}$ is broken regardless since the (pseudo-)periodicity does not corresponds to an odd number of layers anyway. Moreover, the supermodulation leads to a monoclinic distortion, which breaks the $C_{2x}$ symmetry (Extended Data Fig.~\ref{Extended_TEM_300K_100K}\textbf{a} inset). Overall, our TEM data shows that the magnetic point group of UOTe is reduced from $4/mmm1'$ without superlattice to $m_y'$ with superlattice, which is confirmed by our single-crystal XRD measurements (SI.I.3).  We have further performed TEM measurements at 100 K (Extended Data Fig.~\ref{Extended_TEM_300K_100K}\textbf{d}), where chiral imbalance is also observed. We have derived an order parameter to quantify the net chirality based on the amplitude and phase of the TEM displacement, which are indeed nonzero for our TEM data at both 300K and 100K, shown in Table I and II in Methods.3.

To further confirm the net chirality in the superlattice, we also performed optical circular dichroism (CD) \cite{papakostas2003optical, bai2007optical}. At $T>T_{\textrm{N}}$, we only expect the chiral CD. Interestingly, our CD map at $T>T_{\textrm{N}}$ (Figs.~\ref{Fig1}\textbf{j,k}) identified chiral domains. The chiral superlattice reduces the symmetry to a ferromagnetic group. We detect the spontaneous net magnetization $M_z$ by nitrogen-vacancy (NV) center magnetometry: As shown in Fig.~\ref{Fig1}\textbf{m}, the $M_z$ $\sim10\ m\mu_{\textrm{B}}$ per U ion - for comparison, the free moment of each U is $\sim3.2\ \mu_{\textrm{B}}$.

\vspace{0.5cm}
\textbf{Chiral supermodulation spatially encodes Berry curvature}

The chiral superlattice can modulate the low-energy Bloch wavefunction and quantum geometry in real space, generating strong Berry curvature. Since the conduction electrons are composed of the U $6d$ orbitals, we therefore describe the conduction electrons of UOTe by a two-component orbital wavefunction $|u_{\bf r}\rangle=\begin{pmatrix}
C_{\textrm{U1}} \\
C_{\textrm{U2}}
\end{pmatrix}$. This wavefunction is expressed on a Bloch sphere, where the north and south poles correspond to the purely U1 and purely U2 $6d$ states, and the equator corresponds to equal weight superposition. As the conduction electron moves through the superlattice, the Bloch wavefunction rotates strongly on the Bloch sphere between the U1 and U2 (Fig.~\ref{Fig2}\textbf{a}). Not only the Bloch wavefunction, but also the quantum geometry such as Berry curvature is spatially modulated. $x=0$ is polarized to U2, which has left-handed ($z-ix$) chiral displacement. This leads to opposite Berry curvature at $+,-k_y$, therefore a finite Berry curvature dipole along $k_y$ (calculations in Extended Data Fig.~\ref{Extended_Berry_curv_THY_AFM}\textbf{e}). $x=a_s/2$ is polarized to U1, which has right-handed chiral displacement, which in turn leads to the opposite Berry curvature dipole. Due to the net chirality, the opposite Berry curvature dipoles at $x=0$ and $x=a_s/2$ do not cancel. Therefore, the final Berry curvature dipole of the chiral superlattice is nonzero, which can be detected by the nonlinear Hall effect.

We therefore perform experimental studies of the nonlinear Hall (NLH) effect. Figure~\ref{Fig2}\textbf{d} shows the measured nonlinear voltages along both longitudinal and transverse directions. Interestingly, the nonlinear longitudinal signal, $V_{yyy}^{2\omega}$, is nearly zero, whereas the nonlinear Hall signal $V_{xyy}^{2\omega}$ shows a clear quadratic signal. We perform angular dependence.  As shown in Fig.~\ref{Fig2}\textbf{e}, the nonlinear Hall signal is large for $V_{xyy}^{2\omega}$, but zero for $V_{yxx}^{2\omega}$, consistent with the large Berry curvature dipole along $k_y$. We further perform scaling analysis by temperature dependent measurements. The scaling law (Fig.~\ref{Fig2}\textbf{g}) is indeed linear, consistent with the Berry curvature dipole mechanism \cite{sodemann2015quantum}. The nonlinear Hall conductivity is larger than the noncentrosymmetric Weyl semimetal WTe$_2$ \cite{ma2019observation,Kang2019nonlinear} and TaIrTe$_4$ \cite{kumar2021room}, suggesting the potential of the chiral supermodulation as a new and effective way for generating Berry curvature.

\vspace{0.5cm}
\textbf{Chiral superlattice + antiferromagnetic order generates large AHE}

We have shown that the chiral superlattice generates strong Berry curvature via the spatial modulation of quantum geometry. The collinear AFM order can further strongly reshape the $k$ space distribution of the Berry curvatures, breaking their cancellation at $\pm k_y$ (Fig.~\ref{Fig3}\textbf{a}), leading to strong topological magnetic phenomena.

As we go below $T_{\textrm{N}}$, we observe clear AHE that emerges abruptly near $T_\textrm{N}$ (Fig.~\ref{Fig3}\textbf{c}), almost like a step function, which is unusual. The abrupt onset reflects the rapid lifting of $k$-space cancellation of opposite Berry curvature when AFM order sets in. We have performed additional checks to address alternative possibilities including critical scattering and sudden change of carrier mobility (SI.VII.2). Our analysis shows that the change of mobility is mild (maximally $\sim20\%$), therefore unlikely to account for the abrupt onset of AHE near $T_\textrm{N}$.

In order to check that it is the AFM order that couples to the Berry curvature, we have performed systematic AHE and magnetization measurements. This allowed us to show large AHE at zero magnetization at specific B fields (Extended Data Fig.~\ref{Extended_Excluding_M}\textbf{b}). We can further separate the different contributions to the Hall effect (Extended Data Fig.~\ref{Extended_Excluding_M}\textbf{d}), and the intrinsic AFM contribution dominates. We also study other symmetry breaking methods such as magnetic field, interfacial field and strain field, and show that they do not lead to similarly striking low-energy topological phenomena (SI.VII.3). These systematic measurements show that the weak ferromagnetism is a side-product of the symmetry breaking. Moreover, it is important to distinguish this effect from conventional Berry-curvature mechanisms without emergent chiral superlattice. In conventional spin-orbit-coupled chiral AFM metals, Berry curvature is generated by the symmetry breaking from the atomic lattice. By contrast, the atomic lattice of UOTe has $\mathcal{PT}$ and $t\mathcal{T}$ symmetries that forbid the Berry curvature. Therefore, the symmetry breaking should arise beyond the atomic lattice. Our TEM and XRD showed that the only detected symmetry breaking beyond the atomic lattice is the chiral superlattice. Hence, it is the chiral superlattice that breaks the relevant symmetries and strongly modulates the local covalent hybridization, thereby providing the primary route to Berry-curvature generation. Additional measurements and analyses, including within-UOTe control experiments, are discussed in SI.VII.3.1. 

Importantly, the AHE angle ($\sim0.14$) is the largest in AFMs and one of largest in bulk magnets \cite{ye2018massive, li2020giant, sakai2018giant, kim2018large, liu2018giant, nakatsuji2015large,nayak2016large, liu2018electrical, ghimire2018large, feng2022anomalous, gonzalez2023spontaneous, reichlova2024observation, yamada2025metallic}. We further perform scaling analysis based on temperature dependent measurements (Fig.~\ref{Fig3}\textbf{g}). By using $\sigma_{\text{AHE}} = \sigma_{\text{AHE}}^{\text{intrinsic}} + C^{\text{skew}}\sigma_{\text{Drude}}^2$,  we got $\sigma_{\textrm{AHE}}^{\textrm{Intrinsic}}\simeq160$ S/cm or $0.34\ e^2/h$ per layer.

\vspace{0.5cm}
\textbf{Spin-polarized current from collinear antiferromagnet}

Another crucial low-energy magnetic phenomena is the spin-polarized current. Large spin-polarized current in collinear AFMs is highly desired as highlighted by the community of altermagnets and unconventional spin-split magnets \cite{jungwirth2026symmetry, chen2014anomalous, jungwirth2016antiferromagnetic, vsmejkal2018topological,smejkal_beyond_2022,mazin_editorial_2022,cheong2025altermagnetism,song2025altermagnets, nakatsuji2015large} but remains very rare. UOTe is an excellent candidate: The small chiral imbalance combined with the fact that the U $5f$ orbitals are fully filled make the net magnetization small (SI.Fig.S50); On the other hand, the spin-polarized current is only determined by the Fermi surface electrons deriving from the U $6d$ orbitals. Therefore, the spin-polarization of the current can be large in UOTe although the net magnetiation is small.

To directly measure the spin-polarized current generation by UOTe, we fabricated a standard graphene-based spin-transport Hanle device (Figs.~\ref{Fig4}\textbf{a,b}). Figure~\ref{Fig4}\textbf{c} shows the measured Hanle precession of the UOTe-injected spins in graphene. As expected, the detected nonlocal spin signal $R_{\textrm{nl}}(B)$ follows the Hanle profile: the signal is small near $B_x = 0$, then grows as $B_x$ increases. Moreover, the Hanle signals change the polarity and become nearly opposite (blue and red curves in Fig.~\ref{Fig4}\textbf{c}). Subtracting the two curves isolates the pure Hanle precession signal (Fig.~\ref{Fig4}\textbf{d}). In SI.VIII.2 we further show consistent Hanle signal when flipping the magnetization direction of the Co detector. These systematic observations provide evidence for the spin-polarized currents from the collinear AFM UOTe. 
We fit our Hanle data to obtain the spin injection efficiency $\eta_{\textrm{Injection}}$ -- the spin polarization injected into graphene, which provides a lower bound of the Fermi surface spin polarization $S_z^{\varepsilon_{\textrm{F}}}$ of UOTe (see details in Methods.7). We obtain a spin injection efficiency of $\eta_{\textrm{Injection}}=5.40 \pm 0.99 \%$. Additional UOSe/graphene Hanle experiments as well as DFT calculations of the UOTe/graphene proximity effect are discussed in SI.VIII.3. These systematic data and calculations, taken collectively, support the intrinsic spin-polarized current generated by the chiral superlattice over the extrinsic magnetic proximity effect. To further strengthen this conclusion, future Hanle experiment where the chiral superlattice can be in situ switched on and off is desired.

\vspace{0.5cm}
\textbf{Theoretical modeling}

To confirm the chiral superlattice induced Berry curvature and the resulting topological magnetic phenomena, we have built a minimal tight-binding model. In this model, one unit cell contains two U sites, each of which has a $6d$ orbital. The chiral superlattice is then introduced as chiral potential modulation, derived from the chiral superlattice TEM displacement pattern with a periodicity of 7 unit cells. The spatial modulation of the Berry curvature is directly seen from the model in Extended Data Fig.~\ref{Extended_Berry_curv_THY_AFM}\textbf{e}. The resulting superlattice band structure has large Berry curvature with opposite sign at $\pm k_y$ (Extended Data Fig.~\ref{Extended_Berry_curv_THY_AFM}\textbf{b}). We calculate the Berry curvature dipole. Shown in Extended Data Fig.~\ref{Extended_Berry_curv_THY_AFM}\textbf{c}, we found significant $D_{yz}$ around $1-2$ nm near Fermi energy, consistent with the value ($\sim1.62$ nm) derived from our nonlinear Hall data (see SI.IX.4.5). \color{black}  By further including the AFM order into the model, one can see that the AFM order lifts the cancellation between the contributions at $\pm k_y$, thus leading to a net Berry curvature  (Extended Data Figs.~\ref{Extended_Berry_curv_THY_AFM}\textbf{f-j}). The chiral supermodulated AFM band structure is a spin-split semimetal (SI.Fig.S47), facilitating large spin-polarized current  (Figs.~\ref{Fig4}\textbf{e,f} and SI.Fig.S50).  We also used DFT to calculate the chiral superlattice band structure for quantitative comparison with experiments. A reasonable agreement between data and calculation regarding the AHE conductivity is achieved (SI.IX.3.3). Moreover, the large Hall angle ($\sim0.14$) in our data is also captured by our calculations (SI.IX.3.4 and SI.IX.4.4).

\vspace{0.5cm}
\textbf{Formation mechanism and a proposed design principle for chiral supermodulatioin}


Motivated by the strong topological magnetic phenomena observed in UOTe, we uncovered the formation mechanism of the chiral supermodulation. UOTe consists of two distinct covalent bonds, U-O and U-Te. The supermodulation originates from the mismatch between the distinct covalent bonds, which leads to competition between interlayer bonding (energy gain) and intralayer repulsion (energy cost) that drives the supermodulation instability. The mechanism can be visualized by comparing UOS and UOTe (Extended Data Fig.~\ref{Superlattice_Mechanism_1}). UOS has an additional interlayer bond connecting the U with the S in the next layer, which is enabled by the large S-U-S angle ($82$ deg). This interlayer bond lowers energy. However, forming similar interlayer bonding in UOTe would laterally stretch the U-O-U central block due to the much longer U-Te bond length, which is more energy costly. As a result, in the nominal unit cell of UOTe, the Te-U-Te bond angle is much smaller ($77$ deg), which hinders the interlayer bonding. To form the interlayer bonding while keeping the central block un-stretched, the layers decide to locally curve up/down with opposite sign in adjacent layers (Extended Data Fig.~\ref{Superlattice_Mechanism_2}\textbf{a}), which is exactly the supermodulation. In this way, at specific locations, the Te-U-Te angle opens up ($81$ deg), enabling the interlayer U-Te covalent bonding. The cost is that, at some other locations, the Te-U-Te angle reduces ($73$ deg), leading to an intralayer repulsion. This mechanism naturally explains the existence of opposite chirality (SI.II.2). The net chirality can be understood as a follow-up transition due to interlayer bonding anharmonicity (SI.II.2). More broadly, this microscopic mechanism motivates a comparison between two types of superlattices. UOTe is a bond-mismatch superlattice, in which the superlattice strongly modifies the local U-O and U-Te covalent bonding. By contrast, moiré systems are lattice-mismatch superlattices, in which the superlattice modulates the local vdW stacking. The modulation energy scale in UOTe is roughly one order of magnitude stronger than typical moire systems (SI.II.5), making the local quantum geometry a natural description of its physics. Broadly, our data above show that, beyond the artificial moire superlattice, the natural bond-mismatch superlattice can also generate quantum geometrical physics, with a higher energy scale and real-space picture. This provides fresh perspective in designing superlattice materials.

Generally, we can consider three distinct phases: (1) 3D covalent with interlayer bonding, (2) 2D vdW without interlayer bonding, (3) supermodulation as an intermediate phase (Fig.~\ref{Fig5}\textbf{a}). A generic Ginzburg Landau theory can be written as $F=-\alpha\psi^2+\beta\psi^4+p\psi+\lambda\psi\frac{d^2\psi}{dx^2}+\gamma(\frac{d^2\psi}{dx^2})^2 $, where $\psi(x)$ is the interlayer distance. The first two terms provide a standard double-well free energy profile for the 2D vdW and 3D covalent phases, which can be switched by varying the interlayer force $p$. The crucial new term  $\lambda\psi\frac{d^2\psi}{dx^2}$ provides an energy gain when the layer curves; The new term $\gamma(\frac{d^2\psi}{dx^2})^2$ adds an energy penalty for finite curvature. Interestingly, the supermodulation indeed shows up as an intervening phase (Fig.~\ref{Fig5}\textbf{c}). Moreover, our model predicts exotic soliton-superlattices at the boundaries (SI.II.1).  This formation mechanism is verified by calculating the phonon dispersions of the UOX class. (Extended Data Fig.~\ref{Superlattice_Mechanism_2}\textbf{c}).

Based on the above mechanism and phenomenological theory, we propose a design principle that goes from chemistry to physics to materials (Figs.~\ref{Fig5}\textbf{a,d} and Extended Data Fig.~\ref{Design_principle}). From chemistry, we can tune the size of the ions and the strength of chemical bonds. To physics, the above chemical tuning of ion size and bond strength effectively controls the interlayer bonding and intralayer repulsion energies. To materials, there are $\sim500$ isostructural compounds (Extended Data Fig.~\ref{Design_principle}\textbf{c}), generically written as ABC. We have developed an empirical rule based on material parameters such as the ion size and the ratios of bond lengths (Extended Data Fig.~\ref{Design_principle} and SI.II.3). For example, the ratio $r_1=\frac{\textrm{Inter-layer A-C bond length}}{\textrm{Intra-layer A-C bond length}}>1.1$ is an indicator for supermodulation. Based on such design principle, we have theoretically identified promising candidates (see Figs.~\ref{Fig5}\textbf{e-g} and SI.II.4). 

\vspace{0.5cm}
\textbf{Outlooks}

Looking forward, it would be of interest to perform time-domain pump probe experiments where the pump-induced oscillation of the frozen chiral phonons can lead to dynamical control of the observed topological magnetic phenomena. It would be further interesting to study the pressure effect to search for the novel soliton-superlattice phase. More broadly, our work opens up a pathway to similar bond-mismatch superlattices featuring strong modulation energy scale including magnets, superconductors, quantum light sources for new physics, chemistry and device concepts.

\vspace{0.5cm}

\bibliographystyle{naturemag}
\bibliography{AHE_UOTe_2025_v2}

\vspace{0.5cm}
\vspace{0.3cm}
\textbf{Methods}

\textbf{1. Bulk crystal growth:} 

The growth of UOTe was performed in a two-zone furnace, using equimolar ratios of depleted uranium (sourced from New Brunswick Laboratory) and tellurium (99.9999\%, sourced from Alfa Aesar). Growth details are in Ref.~\cite{broyles_uote_2025}.  The source materials were sealed in an evacuated quartz tube, together with $1$ mg/cm$^3$ iodine. The ampoule was place in a two-zone furnace and gradually heated up and held in the temperature gradient of $1030/970$ $^{\circ}$C for 7 days, after which it was cooled to room temperature. The crystal structure was determined by x-ray powder diffraction using a Rigaku x-ray diffractometer with Cu-K$\alpha$ radiation.

\vspace{3mm}
\textbf{2. UOTe Device Fabrication:} 

All fabrication steps involving UOTe were carried out in an argon-filled glovebox without exposure to air, heat, or solvents. For the charge transport devices, thin flakes of UOTe were mechanically exfoliated from bulk crystals onto SiO$_2$/Si substrates.  Electrical contacts (Cr/Au, 5nm/35nm) were deposited via e-beam evaporation through a pre-patterned SiN stencil mask. A protective hexagonal boron nitride (BN) layer (20-50nm thick) was then transferred onto the UOTe flake. For the spin transport devices, monolayer graphene flakes were mechanically exfoliated onto SiO$_2$/Si wafers and identified by optical microscopy. Spin detector contacts (Al$_2$O$_3$/Co/Au) were defined using standard e-beam lithography and deposited via e-beam evaporation. Thin UOTe flakes were exfoliated onto a PDMS stamp and subsequently transferred onto the graphene channel. Hall bar contacts (Cr/Au, 5nm/35nm) were deposited on UOTe using the same stencil mask method described above. A BN flake was transferred onto the final heterostructure to encapsulate and protect both UOTe and the Co contacts from air exposure. For both device types, wire bonding was performed inside the glovebox. The bonded chip carrier was sealed under argon by applying vacuum grease along the rim of the chip carrier and covering it with a glass slide. For the NV center magnetometry measurements, thin UOTe flakes were exfoliated onto a PDMS stamp and transferred onto the diamond substrate.

\vspace{3mm}
\textbf{3. TEM measurements and data analysis:} 

Annular dark-field scanning transmission electron microscopy (STEM) was performed on ThermoFisher Scientific (TFS) Themis Z G3, operated at 200 keV with convergence semi-angle of 18.9 mrad. 20 rapid acquisition frames were registered and averaged to produce drift-free, high signal-to-ratio images. Supermodulation manifests as satellite peaks in Fourier space. Therefore, an image of prisitine, undistorted lattice without periodic lattice displacements can be obtained by damping relevant Fourier components~\cite{savitzky2017}. By comparing atomic coordinates of the original image and the Fourier-damped image, we extract the supermodulation displacement. The atomic coordinates were extracted using Atomap package~\cite{nord2017} (see also SI.I.1.). The precise superlattice in-plane period is determined by diffraction in TEM to be $6.686\pm0.001$.

We further fit the TEM displacement data using the following expression:
\vspace{-0mm}
\begin{align}
    \mathbf{\Delta}_{\textrm{U}1} & =  A_z^{\textrm{U}1}\cos(qx)\hat{z}+A_x^{\textrm{U}1}\sin(qx+\phi_q^{\textrm{U}1})\hat{x}+B_x^{\textrm{U}1}\sin(2qx+\phi_{2q}^{\textrm{U}1}) \nonumber \\
    \mathbf{\Delta}_{\textrm{U}2} & =  A_z^{\textrm{U}2}\cos(qx)\hat{z}-A_x^{\textrm{U}2}\sin(qx+\phi_q^{\textrm{U}2})\hat{x}+B_x^{\textrm{U}2}\sin(2qx+\phi_{2q}^{\textrm{U}2}) \nonumber
\end{align}

The fitted values are shown in Extended Data Fig.~\ref{Extended_TEM_300K_100K} and also the tables below.

\begin{table}[h]
\centering
\scriptsize
\setlength{\tabcolsep}{3pt}
\caption{TEM displacement fitting values at 300~K.}
\begin{tabular}{lccccccccccc}
\hline
Layer & $A_z^{U1}\,(\mathrm{pm})$ & $A_x^{U1}\,(\mathrm{pm})$ & $B_x^{U1}\,(\mathrm{pm})$ & $\phi_q^{U1}\,(\mathrm{deg})$ & $\phi_{2q}^{U1}\,(\mathrm{deg})$ & $A_z^{U2}\,(\mathrm{pm})$ & $A_x^{U2}\,(\mathrm{pm})$ & $B_x^{U2}\,(\mathrm{pm})$ & $\phi_q^{U2}\,(\mathrm{deg})$ & $\phi_{2q}^{U2}\,(\mathrm{deg})$ & $\Psi\,(\mathrm{pm})$ \\
\hline
Layer 1 & $31.7 \pm 0.7$ & $7.8 \pm 0.7$ & $1.5 \pm 0.8$ & $3.5 \pm 5.6$ & $-23.4 \pm 27.8$ & $31.5 \pm 1.1$ & $6.2 \pm 1.1$ & $0.9 \pm 1.1$ & $-1.0 \pm 10.1$ & $-22.1 \pm 69.2$ & 2.0 \\
Layer 2 & $33.7 \pm 0.6$ & $7.9 \pm 0.6$ & $1.6 \pm 0.6$ & $6.4 \pm 4.7$ & $-27.5 \pm 22.3$ & $34.0 \pm 1.0$ & $7.0 \pm 0.9$ & $1.3 \pm 1.0$ & $-9.4 \pm 7.9$ & $-37.3 \pm 42.6$ & 3.1 \\
Layer 3 & $32.8 \pm 0.6$ & $7.7 \pm 0.6$ & $1.6 \pm 0.6$ & $14.4 \pm 4.9$ & $-27.6 \pm 23.8$ & $34.8 \pm 0.8$ & $7.4 \pm 0.8$ & $2.1 \pm 0.8$ & $-12.9 \pm 6.3$ & $-23.6 \pm 21.6$ & 2.1 \\
Layer 4 & $32.8 \pm 0.7$ & $7.4 \pm 0.7$ & $1.6 \pm 0.7$ & $10.2 \pm 5.3$ & $-5.3 \pm 23.9$ & $32.4 \pm 1.1$ & $8.2 \pm 1.1$ & $2.2 \pm 1.1$ & $-8.3 \pm 7.9$ & $-30.8 \pm 28.2$ & 3.1 \\
Layer 5 & $32.6 \pm 0.6$ & $7.2 \pm 0.6$ & $1.7 \pm 0.6$ & $5.1 \pm 5.1$ & $-27.5 \pm 21.0$ & $32.6 \pm 0.9$ & $8.5 \pm 0.9$ & $2.5 \pm 0.9$ & $-18.8 \pm 6.5$ & $-32.5 \pm 21.5$ & 2.9 \\
Layer 6 & $32.0 \pm 0.6$ & $7.6 \pm 0.6$ & $1.8 \pm 0.6$ & $-1.4 \pm 5.0$ & $-32.1 \pm 20.5$ & $33.7 \pm 1.0$ & $8.2 \pm 1.0$ & $2.8 \pm 1.0$ & $-0.5 \pm 7.4$ & $-37.5 \pm 21.5$ & -3.2 \\
Layer 7 & $33.7 \pm 0.8$ & $7.6 \pm 0.8$ & $2.3 \pm 0.8$ & $-1.7 \pm 6.5$ & $-54.9 \pm 20.9$ & $32.1 \pm 1.0$ & $7.7 \pm 1.0$ & $1.9 \pm 0.9$ & $6.8 \pm 7.3$ & $-15.5 \pm 29.9$ & -3.8 \\
Layer 8 & $36.1 \pm 0.9$ & $8.3 \pm 0.9$ & $1.5 \pm 0.9$ & $-1.5 \pm 6.3$ & $-29.5 \pm 34.0$ & $34.2 \pm 0.8$ & $7.5 \pm 0.8$ & $2.0 \pm 0.8$ & $-3.0 \pm 6.6$ & $-27.2 \pm 24.4$ & -3.7 \\
Layer 9 & $36.0 \pm 1.0$ & $9.2 \pm 0.9$ & $2.2 \pm 1.0$ & $4.2 \pm 6.2$ & $-45.2 \pm 25.0$ & $36.7 \pm 1.2$ & $8.5 \pm 1.2$ & $2.1 \pm 1.2$ & $-6.2 \pm 8.1$ & $-19.0 \pm 32.4$ & -2.9 \\
Layer 10 & $34.3 \pm 0.8$ & $8.4 \pm 0.8$ & $2.0 \pm 0.8$ & $8.0 \pm 5.8$ & $-27.8 \pm 24.0$ & $33.6 \pm 1.2$ & $7.9 \pm 1.2$ & $2.5 \pm 1.2$ & $-14.0 \pm 9.1$ & $-30.6 \pm 28.1$ & -1.3 \\
Layer 11 & $34.9 \pm 0.7$ & $8.6 \pm 0.7$ & $1.8 \pm 0.7$ & $13.9 \pm 4.9$ & $-19.5 \pm 22.1$ & $36.6 \pm 1.1$ & $8.5 \pm 1.1$ & $2.9 \pm 1.1$ & $-16.4 \pm 7.5$ & $-34.8 \pm 22.1$ & 3.2 \\
Layer 12 & $33.6 \pm 0.8$ & $7.8 \pm 0.8$ & $2.3 \pm 0.8$ & $8.4 \pm 5.8$ & $-4.7 \pm 19.2$ & $33.5 \pm 1.1$ & $8.6 \pm 1.1$ & $2.1 \pm 1.1$ & $-10.0 \pm 7.6$ & $-48.0 \pm 30.4$ & 4.8 \\
Layer 13 & $34.4 \pm 0.9$ & $8.0 \pm 0.8$ & $2.1 \pm 0.9$ & $6.8 \pm 6.4$ & $-22.1 \pm 23.8$ & $38.4 \pm 1.4$ & $8.4 \pm 1.4$ & $2.1 \pm 1.4$ & $-7.6 \pm 10.0$ & $-33.8 \pm 39.0$ & 3.7 \\
Layer 14 & $32.2 \pm 0.8$ & $8.4 \pm 0.7$ & $1.3 \pm 0.8$ & $9.6 \pm 5.3$ & $-8.1 \pm 32.7$ & $33.1 \pm 1.1$ & $8.8 \pm 1.1$ & $2.1 \pm 1.1$ & $-1.0 \pm 7.6$ & $-10.6 \pm 30.6$ & -3.1 \\
Layer 15 & $31.4 \pm 1.0$ & $8.1 \pm 1.0$ & $2.3 \pm 1.0$ & $6.0 \pm 7.6$ & $-7.9 \pm 26.7$ & $31.0 \pm 1.2$ & $8.2 \pm 1.2$ & $2.4 \pm 1.2$ & $-5.0 \pm 8.7$ & $30.9 \pm 28.6$ & 0.4 \\
\hline
\end{tabular}
\end{table}

\begin{table}[h]
\centering
\scriptsize
\caption{TEM displacement fitting values at 100~K.}
\setlength{\tabcolsep}{3pt}
\begin{tabular}{lccccccccccc}
\hline
Layer & $A_z^{U1}\,(\mathrm{pm})$ & $A_x^{U1}\,(\mathrm{pm})$ & $B_x^{U1}\,(\mathrm{pm})$ & $\phi_q^{U1}\,(\mathrm{deg})$ & $\phi_{2q}^{U1}\,(\mathrm{deg})$ & $A_z^{U2}\,(\mathrm{pm})$ & $A_x^{U2}\,(\mathrm{pm})$ & $B_x^{U2}\,(\mathrm{pm})$ & $\phi_q^{U2}\,(\mathrm{deg})$ & $\phi_{2q}^{U2}\,(\mathrm{deg})$ & $\Psi\,(\mathrm{pm})$ \\
\hline
Layer 1 & $32.2 \pm 1.2$ & $5.1 \pm 1.1$ & $1.9 \pm 1.2$ & $29.1 \pm 13.4$ & $-24.2 \pm 34.3$ & $31.3 \pm 0.9$ & $10.0 \pm 0.9$ & $0.2 \pm 0.9$ & $-4.4 \pm 5.2$ & $35.4 \pm 292.0$ & 5.7 \\
Layer 2 & $32.1 \pm 1.3$ & $5.7 \pm 1.2$ & $2.9 \pm 1.2$ & $16.7 \pm 12.7$ & $-2.7 \pm 24.9$ & $31.9 \pm 1.0$ & $9.3 \pm 1.0$ & $2.1 \pm 1.0$ & $-0.5 \pm 6.6$ & $-34.3 \pm 28.2$ & 6.1 \\
Layer 3 & $34.0 \pm 1.3$ & $5.4 \pm 1.2$ & $2.5 \pm 1.2$ & $18.3 \pm 13.4$ & $-21.9 \pm 28.8$ & $32.6 \pm 1.1$ & $9.3 \pm 1.1$ & $2.2 \pm 1.1$ & $4.3 \pm 6.9$ & $-15.3 \pm 28.5$ & 5.3 \\
Layer 4 & $34.2 \pm 0.8$ & $7.2 \pm 0.8$ & $2.7 \pm 0.8$ & $20.4 \pm 6.6$ & $2.7 \pm 17.7$ & $34.1 \pm 0.9$ & $8.0 \pm 0.9$ & $1.6 \pm 0.9$ & $-0.3 \pm 6.5$ & $-69.3 \pm 31.1$ & 5.9 \\
Layer 5 & $34.7 \pm 1.3$ & $7.6 \pm 1.3$ & $2.2 \pm 1.3$ & $15.0 \pm 10.0$ & $-18.1 \pm 33.2$ & $34.6 \pm 1.2$ & $9.4 \pm 1.2$ & $1.9 \pm 1.2$ & $-3.8 \pm 7.6$ & $-28.2 \pm 35.8$ & 5.2 \\
Layer 6 & $34.3 \pm 1.2$ & $5.7 \pm 1.2$ & $3.1 \pm 1.1$ & $14.9 \pm 12.1$ & $-13.5 \pm 22.5$ & $33.2 \pm 1.0$ & $9.4 \pm 0.9$ & $1.4 \pm 1.0$ & $-4.9 \pm 6.0$ & $-39.0 \pm 38.3$ & 6.4 \\
Layer 7 & $33.7 \pm 1.2$ & $6.9 \pm 1.2$ & $1.3 \pm 1.2$ & $9.0 \pm 10.3$ & $-38.5 \pm 54.3$ & $33.3 \pm 1.1$ & $10.2 \pm 1.1$ & $1.7 \pm 1.1$ & $-8.0 \pm 6.5$ & $-12.8 \pm 37.4$ & 3.8 \\
Layer 8 & $32.7 \pm 1.2$ & $6.4 \pm 1.2$ & $1.2 \pm 1.2$ & $13.0 \pm 11.2$ & $-5.7 \pm 59.3$ & $30.7 \pm 1.0$ & $9.3 \pm 1.0$ & $1.1 \pm 1.0$ & $-4.9 \pm 6.3$ & $7.5 \pm 49.1$ & 3.9 \\
Layer 9 & $31.3 \pm 1.5$ & $6.3 \pm 1.5$ & $2.5 \pm 1.5$ & $-2.0 \pm 14.3$ & $-30.9 \pm 35.3$ & $31.2 \pm 0.9$ & $9.3 \pm 0.9$ & $1.2 \pm 0.9$ & $4.9 \pm 5.8$ & $9.2 \pm 42.3$ & 5.0 \\
Layer 10 & $34.6 \pm 1.5$ & $5.1 \pm 1.4$ & $2.1 \pm 1.4$ & $17.0 \pm 16.5$ & $7.7 \pm 39.3$ & $34.2 \pm 1.3$ & $9.6 \pm 1.2$ & $1.4 \pm 1.2$ & $-1.9 \pm 7.8$ & $41.0 \pm 53.4$ & 5.4 \\
Layer 11 & $35.4 \pm 1.5$ & $7.2 \pm 1.5$ & $2.3 \pm 1.5$ & $11.3 \pm 12.3$ & $12.6 \pm 38.1$ & $34.8 \pm 0.9$ & $8.9 \pm 0.9$ & $1.1 \pm 0.9$ & $1.3 \pm 6.0$ & $-6.9 \pm 47.9$ & 5.1 \\
Layer 12 & $37.0 \pm 1.6$ & $6.1 \pm 1.5$ & $3.1 \pm 1.5$ & $5.4 \pm 15.0$ & $16.0 \pm 29.2$ & $33.7 \pm 1.0$ & $10.5 \pm 0.9$ & $1.9 \pm 0.9$ & $-5.4 \pm 5.4$ & $14.7 \pm 28.9$ & 6.2 \\
Layer 13 & $37.1 \pm 1.5$ & $7.8 \pm 1.5$ & $2.4 \pm 1.5$ & $1.6 \pm 11.4$ & $-25.4 \pm 36.1$ & $33.9 \pm 1.0$ & $10.2 \pm 1.0$ & $1.0 \pm 1.0$ & $4.1 \pm 5.9$ & $30.9 \pm 56.3$ & 5.4 \\
Layer 14 & $34.6 \pm 1.7$ & $5.1 \pm 1.7$ & $2.3 \pm 1.7$ & $-0.7 \pm 19.3$ & $-23.5 \pm 43.0$ & $35.5 \pm 1.3$ & $10.3 \pm 1.3$ & $1.9 \pm 1.3$ & $-4.3 \pm 7.5$ & $-16.9 \pm 39.0$ & 5.7 \\
Layer 15 & $32.6 \pm 1.8$ & $5.8 \pm 1.7$ & $2.7 \pm 1.8$ & $8.7 \pm 17.8$ & $-10.4 \pm 37.8$ & $34.3 \pm 0.7$ & $9.4 \pm 0.7$ & $0.4 \pm 0.7$ & $-1.2 \pm 4.3$ & $63.1 \pm 97.8$ & 6.1 \\
\hline
\end{tabular}
\end{table}

Based on the above displacements, one can mathematically derive an order parameter $\Psi$ that breaks both inversion symmetry and mirror symmetry - the net chirality (see SI.IX.1 for derivation)

\begin{equation}
\Psi^3 =- (A_x^{\textrm{U}1})^2 B_x^{\textrm{U}2} \cos\left(\phi_{2q}^{\textrm{U}2}-2\phi_q^{\textrm{U}1}\right)+(A_x^{\textrm{U}2})^2 B_x^{\textrm{U}1} \cos\left(\phi_{2q}^{\textrm{U}1}-2\phi_q^{\textrm{U}2}\right) .
\end{equation}

From the expression of the order parameter above, we can see that the system has NO net chirality, when and only when $A_x^{\textrm{U}1}=A_x^{\textrm{U}2}$, $B_x^{\textrm{U}1}=B_x^{\textrm{U}2}$, $\phi_q^{\textrm{U}1}=-\phi_q^{\textrm{U}2}$ and $\phi_{2q}^{\textrm{U}1}=-\phi_{2q}^{\textrm{U}2}$. No net chirality corresponds to $\Psi=0$. The system has net chirality when the above conditions are not satisfied. The above condition shows that our TEM data has net chirality. Our $300$ K data (Extended Data Fig.~\ref{Extended_TEM_300K_100K}) shows that $\phi_{2q}^{\textrm{U}1}$ and $\phi_{2q}^{\textrm{U}2}$ have the same sign. Our 100 K data (Extended Data Fig.~\ref{Extended_TEM_300K_100K}) shows that not only the U1 and U2 phases have the same sign but also the amplitudes $A_x^{\textrm{U}1}$ and $A_x^{\textrm{U}2}$ show observable differences. In Extended Data Fig.~\ref{Extended_Structural_Simulations}, we provide simulations and intuition for the different terms (amplitudes and phases for first and second harmonics) and how they will lead to net chirality. We can see that Case 1 if we only turn on the first harmonic $z$ and first harmonic $x$ modulations with opposite phases, the $x+iz$ (orange) and $x-iz$ (green) modulations are perfectly symmetric with respect to the inversion center. Case 2 if we now add second-harmonic x modulation with opposite phases, the $x+iz$ (orange) and $x-iz$ (green) modulations are perfectly symmetric with respect to the inversion center. Case 3, if we now change the second harmonic modulation of U1 and U2 to have the same phase, the $x+iz$ (orange) and $x-iz$ (green) modulations are not symmetric any more.

\vspace{3mm}
\textbf{4. Optical measurements:}

\underline{4.1. Linear dichroism:} The superlattice breaks the $C_{4z}$ symmetry, enabling a nonzero optical LD. LD signal can be used to distinguish the orthogonal superlattice domains, $\textbf{q}\parallel a$ and $\textbf{q} \parallel b$.  

\underline{4.2. Circular dichroism:} CD can arise from both chirality and net magnetization. The CD measurements were performed in reflection mode with oblique incidence, in order to avoid the cancellation of the chirality-induced CD signal (the optical activity) at normal incidence due to time-reversal symmetry \cite{fried2014, armitage2014, hosur2015}. At $T>T_{\textrm{N}}$, CD only arises from chirality; the corresponding optical activity tensor is $\sigma_{xyy}$, which is indeed nonzero in the group $m_y'$. At $T<T_{\textrm{N}}$, additional contribution from magnetization arises; However, we can use the magnetic field hysteresis measurements to separate them (see SI.III.3). The magnetic CD is antisymmetric with the magnetic field (the hysteresis loop part) whereas the chiral CD is symmetric with magnetic field (a nonzero offset).

\underline{4.3. Experimental methods:} Optical LD and CD measurements were performed in the magneto-optical cryostat OptiCool by Quantum Design using a supercontinuum laser SuperK Extreme by NKT photonics. A spectrometer SpectraPro-300i by Acton Research was used to select the wavelength. The beam went through a photoelastic modulator. The retardation was half-wave for LD and quarter-wave for CD. Oblique incidence was achieved by sending the laser beam through one side of the objective. The oblique angle was $\sim10^{\circ}$ maximally. The reflected beam went through the cryostat's top window, was collimated by the same objective, and collected by an avalanche photodiode.

\vspace{3mm}
\textbf{5. NV center magnetometry measurement:} We used a $(100)$-oriented electronic grade CVD-grown diamond (Element Six), mechanically thinned down to 50 $\mu$m and polished. A shallow ensemble of nitrogen-vacancy (NV) centers were created by $^{14}$N ion implantation. NV center magnetometry, as reported in previous works on the same setup \cite{andersen2019}, was performed based on the standard optically detected magnetic resonance. We used a solid-state green laser $532$ nm, $\sim$ 1 mW modulated by an acousto-optic modulator (AOM). The control microwave pulses are generated via IQ mixing, using a microwave generator and analog control pulses produced by a high-sampling-rate arbitrary waveform generator (AWG). The signal was amplified by a medium-gain (Mini-Circuits ZX60-6013E-S+) and a high-gain, high power amplifier (Mini-Circuits ZHL-25W-63+). Timing of all optical pulses, AWG trigger, and data acquisition, is orchestrated by a high speed programmable pulse generator (SpinCore PulseBlasterESR-PRO-USB-RM2). A flake of UOTe and a flake of Fe$_3$GeTe$_2$ were both transferred onto the same diamond chip. We performed NV center magnetometry on both flakes (see data in SI.IV)

\vspace{3mm}
\textbf{6. Charge transport:} 

\underline{6.1. Measuring the linear charge anisotropy and AHE:} Consider a crystal with both anisotropy and AHE. The linear resistivity can be written as $\left(\begin{smallmatrix}         \rho_{xx} & \rho_{xy} \\        \rho_{yx} & \rho_{yy}     \end{smallmatrix}\right)=\left(\begin{smallmatrix}           \rho_{aa}\cos^2(\theta)+ \rho_{bb}\sin^2(\theta) & (\rho_{aa}-\rho_{bb})\cos(\theta)\sin(\theta) - \rho_{\textrm{AHE}}\\
        (\rho_{aa}-\rho_{bb})\cos(\theta)\sin(\theta) + \rho_{\textrm{AHE}} & \rho_{aa}\sin^2(\theta)+ \rho_{bb}\cos^2(\theta)    \end{smallmatrix}\right)$, where $a,b$ are the principal crystalline directions, $x,y$ are the lab coordinates and x is the direction of the applied bias, and $\theta$ is the angle between $a$ and $x$. Based on this resistivity tensor, we explain how we measured the charge anisotropy and AHE in UOTe. The longitudinal resistance $\rho_{ii}$ can measure the charge anisotropy, i.e., $\rho_{aa}\neq\rho_{bb}$ (SI.VI.1.1). The transverse resistance $\rho_{yx}=\rho_{yx}^{\textrm{aniso.}}+\rho_{\text{AHE}}$ contains both anisotropy and AHE contributions. $\rho_{yx}^{\textrm{aniso.}}= (\rho_{aa}-\rho_{bb})\cos(\theta)\sin(\theta)$ has a $d$-wave angular pattern and is symmetric with $B$ field (SI.VI.1.2). $\rho_{\textrm{AHE}}$ is angular-independent and and is antisymmetric with $B$ field (Figs.~\ref{Fig3}\textbf{d,e}).  

The conductivity tensor was then obtained by inverting the resistivity tensor  $\left(\begin{smallmatrix}         \sigma_{xx} & \sigma_{xy} \\        \sigma_{yx} & \sigma_{yy}      \end{smallmatrix} \right) = \frac{1}{\rho_{xx}\rho_{yy}-\rho_{xy}\rho_{yx}}\left(\begin{smallmatrix}         \rho_{yy} & -\rho_{yx} \\        -\rho_{xy} & \rho_{xx}      \end{smallmatrix} \right)$. The AHE conductivity was then obtained by the antisymmetrizing the off diagonal conductivity $\sigma_{\textrm{AHE}}=\frac{\sigma_{yx}-\sigma_{xy}}{2}$.

\underline{6.2. Measuring the nonlinear Hall conductivity:} The nonlinear Hall conductivity can be obtained from the injected current $I^{\omega}$ along $y$, the measured nonlinear Hall voltage $V_{xyy}^{2\omega}$, as well as the linear resistances, by $\sigma^{2\omega}_{xyy}=\frac{V^{2\omega}_{xyy}}{{I^{\omega}_{y}}^{2}R_{yy}^2R_{xx}}\frac{l^3}{w^2d}$

\underline{6.3. Experimental methods:} Transport measurements were performed in a Quantum Design Dynacool PPMS system with a base temperature of 1.7K and a maximum magnetic field of 9T. For electrical transport measurements, longitudinal and transverse voltages were recorded simultaneously using standard low-frequency lock-in techniques (SR830 lock-in amplifiers), with the magnetic field applied along the out-of-plane direction.

\vspace{3mm}
\textbf{7. Spin transport:} We fabricated a spin transport device using a standard nonlocal geometry and performed Hanle spin precession measurements in both UOTe-Co configuration (Fig.~\ref{Fig4}) and Co-Co configuration (SI.VIII.2.1). The same monolayer graphene serves as the spin transport channel, while either UOTe or Co contact is used as the spin injector and another Co contact as the spin detector. 

In order to extract the spin polarization of spin injector and detector, we fit the UOTe-Co and the Co-Co Hanle signal curves to the corresponding solutions of the time-independent Bloch equations for a uniform spin diffusion system (see SI.VIII.2). Specifically, we make the following assumption: First, multiple Co contacts have the same spin polarization. Second, In UOTe-Co experiment, we only consider the spin injection at one edge of UOTe flake (closest to the Co detector contact). In this case, we neglect the spin diffusion and relaxation underneath UOTe. This also contributes to the fact that the extracted values provide a lower bound for the spin polarization of the UOTe-covered region. Under these assumptions, we fit the Co-Co Hanle curve to

\vspace{-7mm}
\begin{align}
	R_{\textrm{nl}} = \Big( \eta_{\textrm{Injection}}\Big) * \left( \frac{R_{sq}\lambda_{\mathrm{spin}} \cos{\varphi_1}\cos{\varphi_2}}{2W_1} \frac{e^{-\frac{L_1}{\lambda_{\mathrm{spin}}}\sqrt{1-i\gamma_{\mathrm{e}} B \tau }} + e^{-\frac{L_1}{\lambda_{\mathrm{spin}}}\sqrt{1+i\gamma_{\mathrm{e}} B \tau }} }{\sqrt{1-i\gamma_{\mathrm{e}} B \tau} + \sqrt{1+i\gamma_{\mathrm{e}} B \tau}}\right)*\Big(\eta_{\textrm{Detection}}\Big), \nonumber
\end{align}  

The three brackets represent the three key parts of the Hanle experiment. $\eta_{\textrm{Injection}}$ and $\eta_{\textrm{Detection}}$, which both range from 0\% to 100\%, are the spin injection and detection efficiencies. Because both the injector and detector are Co electrodes prepared in the identical recipe, we assume that $\eta_{\textrm{Injection}}=\eta_{\textrm{Detection}}=\eta_{\textrm{Co}}$. The middle bracket characterizes spin propagation and precession, where $\gamma_{\mathrm{e}}$ is the electron gyromagnetic ratio, $R_{\textrm{sq}}=3.8$ kOhm is the graphene sheet resistance, $L_1=0.80$ $\mu$m and $W_1=4.70$ $\mu$m are the channel length and width, $\lambda_{\mathrm{spin}}$ is the spin relaxation length, $\tau$ is the spin lifetime, and $\varphi_{1,2}$ is angle between the Co contact magnetization and its easy axis. By fitting the data in SI.VIII.2.1, we get $\lambda_{\mathrm{spin}}=2.17\pm0.46$ $\mu$m, $\tau=235.66\pm83.04$ ps, and $\eta_{\textrm{Co}}=1.03\pm0.11\%$.

We can proceed to fit the UOTe-Co Hanle curve by 

\vspace{-7mm}
\begin{align}
	R_{\textrm{nl}} = \Big( \eta_{\textrm{Injection}}\Big) * \left(\frac{R_{sq}\lambda_{\mathrm{spin}} \cos{\varphi_1}}{2W_2} \frac{e^{-\frac{L_2}{\lambda_{\mathrm{spin}}}\sqrt{1-i\gamma_{\mathrm{e}} B \tau }} - e^{-\frac{L_2}{\lambda_{\mathrm{spin}}}\sqrt{1+i\gamma_{\mathrm{e}} B \tau }}}{i(\sqrt{1-i\gamma_{\mathrm{e}} B \tau} + \sqrt{1+i\gamma_{\mathrm{e}} B \tau} )}\right)*\Big(\eta_{\textrm{Detection}}\Big), \nonumber	
\end{align}  
where the $\eta_{\textrm{Injection}}$ is the injection efficiency from UOTe, $\eta_{\textrm{Detection}}$ is the detection efficiency by Co, which has been determined above, $R_{\textrm{sq}}=3.8$kOhm is the graphene sheet resistance, $L_2=1.58$ $\mu$m and $W_2=4.53$ $\mu$m are the channel length and width. Note that the two equations are different because of the different injected spin direction (UOTe injects $S_z$ while Co injects $S_y$). By fitting the data in Fig.~\ref{Fig4}\textbf{d}, we get $\lambda_{\mathrm{spin}}=1.02\pm0.15$ $\mu$m, $\tau=49.76\pm14.68$ ps, and $\eta_{\textrm{UOTe}}=5.40\pm0.99\%$.

\vspace{3mm}
\textbf{8. Tight-binding modeling:} In this model, a unit cell contains two U sites, each of which has a $6d$ orbital, compatible with space group \#129. Therefore, our minimal model has two bands. The Hamiltonian reads $H = - \sum_{\langle i,j \rangle} t_0 \left( a_i^\dagger b_j + b_j^\dagger a_i \right)$. The chiral superlattice is then introduced as a slowly varying chiral potential modulation, derived from the chiral displacement pattern. The superlattice Hamiltonian reads
$H_{\textrm{s}} = - \sum_{\langle i,j \rangle} t(x) \left( a_i^\dagger b_j + b_j^\dagger a_i \right)
+ \sum_i V_1(x)\, a_i^\dagger a_i
+ \sum_j V_2(x)\, b_j^\dagger b_j$. If there is no chiral imbalance, then the U1 $z+ix$ and U2 $z-ix$ are related by the inversion symmetry and $C_2$ screw axis, therefore giving rise to a Dirac nodal crossing.  The chiral imbalance then breaks degeneracy between U1 $z+ix$ and U2 $z-ix$, opening up a gap at the nodal crossings. The resulting superlattice band structure has strong Berry curvature with opposite sign at $\pm k_y$ (Extended Data Fig.~\ref{Extended_Berry_curv_THY_AFM}\textbf{b}), which is generated by the spatial modulation mechanism explained above. The spatial modulation of the Berry curvature $\Omega(x)$ is calculated by treating the slowly varying superlattice modulation as an adiabatic parameter and approximating each local region by a homogeneous tight-binding Hamiltonian $H(\mathbf{k};x)$. The result is shown in Extended Data Fig.~\ref{Extended_Berry_curv_THY_AFM}\textbf{e}, consistent with our intuitive picture in Fig. 2.

We can further include the AFM order. The supermodulated AFM Hamiltonian reads: $H=H_{\textrm{s}}+ H_{\mathrm{SOC}} + H_{\mathrm{AFM}}$, where $H_{\mathrm{AFM}} = m \sum_i \left( a_{i\uparrow}^\dagger a_{i\uparrow} - a_{i\downarrow}^\dagger a_{i\downarrow} \right)
- m \sum_j \left( b_{j\uparrow}^\dagger b_{j\uparrow} - b_{j\downarrow}^\dagger b_{j\downarrow} \right)$ and $H_{\mathrm{SOC}} = -\lambda_{\mathrm{SOC}} \sum_{\langle i,j \rangle}
\begin{pmatrix} a_{i\uparrow}^\dagger & a_{i\downarrow}^\dagger \end{pmatrix}
(\hat{\mathbf{n}}_{ij} \cdot \boldsymbol{\sigma})
\begin{pmatrix} b_{j\uparrow} \\ b_{j\downarrow} \end{pmatrix}
+ \mathrm{h.c.},
$.  The AFM couples strongly to the Berry curvature through the spin-orbit coupling (SOC), lifting the cancellation between $\pm k_y$ and leading to a net Berry curvature in $k$ space, explaining the AHE. The chiral supermodulated AFM band structure is a spin Zeeman split semimetal. With such a low-energy band structure, the spin-polarization of the current will be naturally large.

\vspace{3mm}
\textbf{9. First-principle calculations:} We used first-principle calculations within the DFT framework as implemented in the Vienna {\it ab-initio} simulation package (VASP) \cite{Kresse1996}. The ground state electronic structure was calculated using the projector augmented-wave pseudopotential, while electron exchange-correlation effects are treated with the Local Density Approximation. SOC was incorporated through $j$-dependent pseudopotentials. The energy cutoff of 450 eV was used for the plane-wave basis set, and the BZ integration was performed with a $5\times 9 \times 5$ $k$-mesh. The total energy tolerance criterion is set to $10^{-6}$ eV. Onsite Coulomb potential was $U=4$ so that the energy of the U $5f$ band matches the ARPES data (i.e., $\simeq 2$ eV below the Fermi level) \cite{broyles_uote_2025}. For the atomic lattice, we used the experimentally determined lattice parameter in Ref. \cite{broyles_uote_2025}. For the superlattice lattice structure, we used TEM fitted displacement (see expressions in Methods.3), only changing the periodicity to $7a$ (commensurate). Therefore, the level of chiral imbalance should serve as a reasonable reflection of the actually measured chiral imbalance by TEM. The DFT calculated a magnetic group $m_y'$, which agrees with our TEM and XRD experimental results above.

\vspace{0.5cm}

\textbf{Author contributions:}  TD fabricated the devices. SHS performed the TEM measurements with help from ML, TD and IEB, inspired by early STM measurements of ML and JEH. JXQ initiated the optical measurements and performed optical LD/CD experiments with help from TD and XZ. TD performed the charge transport measurements with help from CZ, ML, YFL and AG. TSG conceived and designed the spin-transport and Hanle experiments, and supervised the measurements performed by TD. XHL performed the NV center magnetometry measurements with the help from TD, MDL, HP. JMB, John Singleton, and RM performed the high-field magnetization measurements at LANL with the help from TD. IA performed the MPMS measurements with help from TD and DC. SLZ performed the single-crystal x-ray diffraction measurements with help from TD. CB, QX, ZR, HC, WX, and SR grew the UOTe single crystals. KW and TT grew the bulk hBN single crystals. SM performed first-principles calculations with help from BG, VKS, SG, Hsin Lin, TRC, TYL, AB, and SC. KS performed effective modeling with help from Jairo Sinova and AV. MS, Jingtian Shi and IM proposed the free-energy model for the chiral superlattice. HCL, PG, JL, TH, CF, QM helped with the charge transport experiments. TD, ML, JXQ, PK and SYX wrote the manuscript with contributions from others. SYX, PK, and ML conceived the experiments. SYX and PK supervised the project.

\textbf{Acknowledgement:} We gratefully thank Sang-Wook Cheong, Aravind Devarakonda, Takashi Kurumaji, Igor Mazin, and Linda Ye for discussions.

\textbf{Funding statement:} TD, JXQ, MS, JMB, RM, QM, Jingtian Shi, IM, AV and SYX were supported through the Center for the Advancement of Topological Semimetals (CATS), an Energy Frontier Research Center (EFRC) funded by the US Department of Energy (DOE) Office of Science, through the Ames National Laboratory under contract DE-AC0207CH11358.  The work in SYX group was supported by the Office of Naval Research (ONR) grant N000142512285 (charge and spin transport), the National Science Foundation (NSF) Career Grant No. DMR-2143177 (fabrication), CATS (data analysis) and the Army Research Office (ARO) under W911NF2420195 Cooperative Agreement Number W911NF-24-2-0195 (manuscript writing). SYX acknowledges the Alfred P. Sloan foundation, the Camille and Henry Dreyfus foundation, and Corning Fund for Faculty Development. PK acknowledges support from AFOSR grant no. FA9550-25-1-0019. TD acknowledges a partial support from AFOSR MURI grant no. FA9550-25-1-0262 (fabrication in the Kim group). QM acknowledges support from the National Science Foundation (NSF) under award No. DMREF-2522383. HP and MDL were supported by the GBMF grant no. 7797-01 and NSF CUA (PHY-1125846 for HP and MDL). HP, MDL, PK and SYX are supported by the Samsung Electronics Co., Ltd award \#A61703. TSG was supported by the Dutch Research Council (NWO) through a Rubicon grant (No. 019.222EN.013) and a Veni grant (No. VI.Veni.242.281). ML acknowledges the Harvard Quantum Initiative Postdoctoral Fellowship and the start-up funds at University of Texas at Dallas. JEH was supported by the Gordon and Betty Moore Foundation (GBMF)'s EPiQS Initiative through grant GBMF10215. I.E.B. and S.H.S. acknowledge support by the Rowland Institute at Harvard. Research at Washington University was supported by the U.S. DOE, Office of Science, Basic Energy Sciences Grant No. DE-SC0026267 (sample synthesis) and National Science Foundation (NSF) Division of Materials Research Award DMR-2236528 (high magnetic field measurement). Electron microscopy experiments were performed at the Michigan Center for Materials Characterization ((MC)\textsuperscript{2}, RRID: SCR\_026770) and MIT.nano. Focused ion beam samples were prepared at (MC)\textsuperscript{2} as well as at the Harvard University Center for Nanoscale Systems (CNS), a member of the National Nanotechnology Coordinated Infrastructure Network (NNCI), which is supported by the National Science Foundation under NSF award No.~ECCS-2025158. A portion of this work was performed at the National High Magnetic Field Laboratory, which is supported by the National Science Foundation Cooperative Agreement No. DMR-2128556, the US Department of Energy (DoE) and the State of Florida. KW and TT were supported by the JSPS KAKENHI (21H05233 and 23H02052) , the CREST (JPMJCR24A5), JST and World Premier International Research Center Initiative (WPI), MEXT, Japan.  CF was supported by Deutsche Forschungsgemeinschaft through SFB 1143 (project ID. 24731007) and the W\"urzburg-Dresden Cluster of Excellence ctd.qmat - Complexity, Topology and Dynamics in Quantum Matter (EXC 2147, project-id 390858490).  HL was supported by the National Science and Technology Council (NSTC) in Taiwan under grant number NSTC 114-2112-M-001-055-MY3. TRC was supported by National Science and Technology Council (NSTC) in Taiwan (NSTC 114-2628-M-006-005-MY3 and NSTC113-2124-M-006-009-MY3), National Cheng Kung University (NCKU), Taiwan, and National Center for Theoretical Sciences, Taiwan. This research was supported, in part, by the Higher Education Sprout Project, Ministry of Education to the Headquarters of University Advancement at NCKU. TRC thanks the National Center for High performance Computing (NCHC) of National Applied Research Laboratories in Taiwan for computational and storage resources. The work at S. N. Bose National Centre for Basic Sciences (SNBNCBS) was supported by Prime Minister Early Career Research Grant (PM-ECRG) from Anusandhan National Research Foundation (ANRF), file number ANRF/ECRG/2024/003677/PMS, and also benefited from the PARAM-Rudra computational facility at SNBNCBS. The work at Howard University was supported by the National Science Foundation through the U.S. Department of Energy (DOE), Office of Science, Basic Energy Sciences, under Grant No. DE-SC0022216 and utilized computational resources provided through ACCESS allocation PHY260045. The work at Northeastern University was supported by the National Science Foundation through the Expand-QISE award NSF-OMA-2329067 and benefited from the resources of Northeastern University Advanced Scientific Computation Center, the Explorer Cluster, the Massachusetts Technology Collaborative award MTC-22032, and the Quantum Materials and Sensing Institute (QMSI). Jairo Sinova was supported by the U.S. DOE, Office of Science, under Award No. DE-SC0026038. We acknowledge the support to the X-ray core facility from the Major Research Instrumentation Program of the NSF under Award Number 2216066. KS was supported by the GBMF10694.

\textbf{Competing financial interests:} The authors declare no competing financial interests.

\clearpage
\addtocounter{figure}{-1}
\begin{figure*}[t!]
\caption{Such supermodulations of U1 and U2 can be viewed as the ``freezing'' of opposite chiral phonons ($z+ix$ for U1 and $z-ix$ for U2). \textbf{f,} Zoomed-in view of the DFT calculated phonon dispersion of UOTe. See full dispersion in Extended Data Fig.~\ref{Extended_DFT_phonon}. The negative phonon frequency at finite $\vec{q}=(\frac{1}{6.7}\frac{2\pi}{a}, 0, \frac{1}{2}\frac{2\pi}{c})$ is consistent with the experimentally observed $\vec{q}$ for the superlattice. \textbf{g,} Schematic illustrations for the atomic displacements for U1 and U2, which have opposite chiral winding structures. The opposite winding structures have slightly different amplitudes and therefore do not cancel each other. Such a spontaneous chiral imbalance leads to a net chirality which breaks $\mathcal{PT}$ symmetry. \textbf{h,i,} Measured out-of-plane and in-plane atomic displacements for U1 and U2. The in-plane displacement $\Delta_{\textrm{U}1 ,x}$ and $\Delta_{\textrm{U}2 ,x}$ have visibly different shapes, which confirm that the opposite chiral displacements of U1 and U2 do not cancel. Circles: TEM data points; Lines: Fitting of the TEM data. The fitting parameter values are shown as Layer 11 in the 300K table in Methods 3. \textbf{j,k,} Optical circular dichroism (CD) setup and data at $T=160$ K. The observation of chiral CD and chiral domains provide further evidence for the net chirality.   \textbf{l,m,} Diamond nitrogen-vacancy (NV) center magnetometry setup and net magnetization map of a UOTe flake measured by NV magnetometry at $100$ K.}
\end{figure*}
\begin{figure*}[t]
\includegraphics[width=14cm]{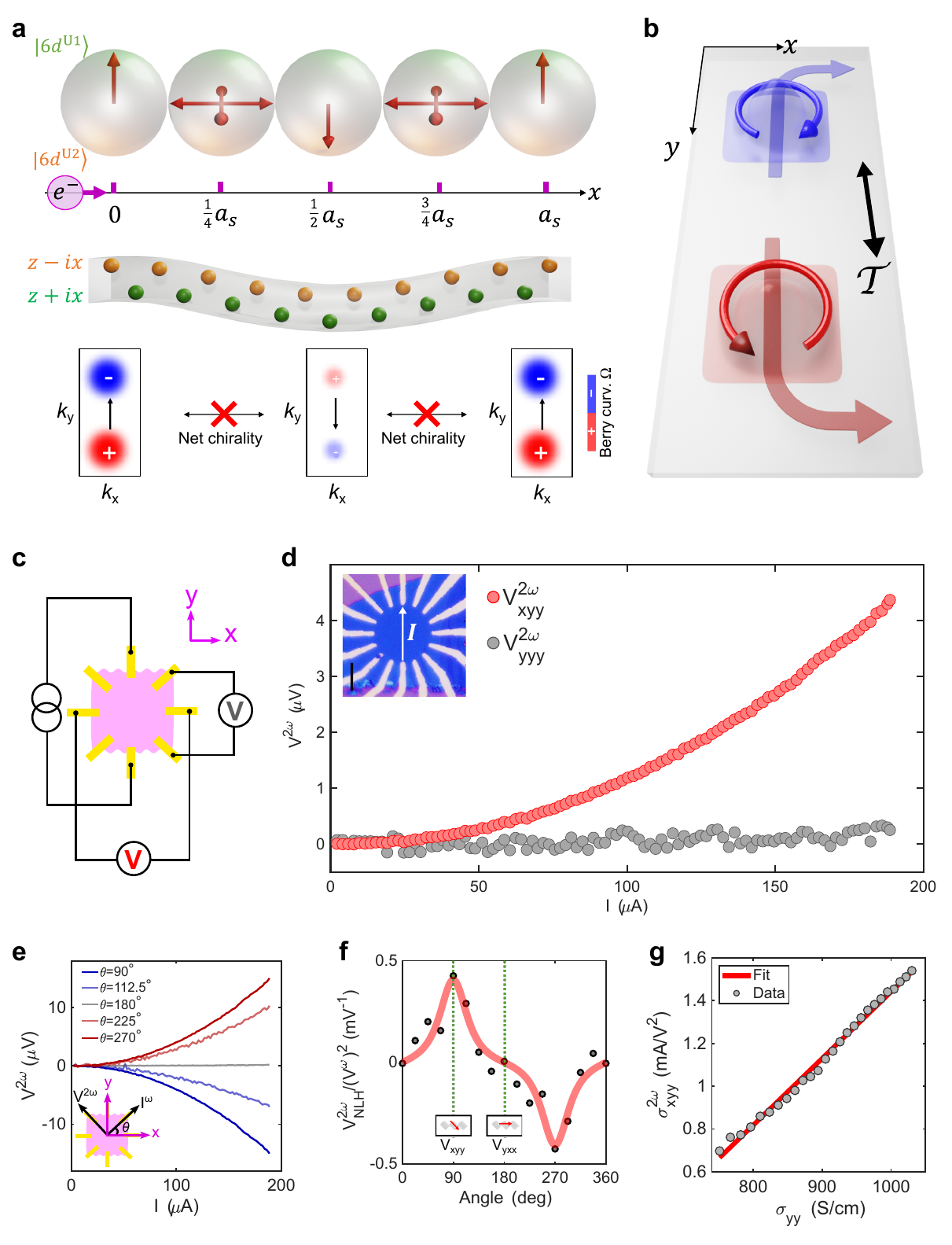}
\vspace{-0.5cm}
\caption{\textbf{Chiral supermodulation generates strong Berry curvature.} \textbf{a,} The conduction electrons composed of U $6d$ orbitals can be described by a two-component wavefunction. As the conduction electron moves through the superlattice, both the Bloch wavefunction and the Berry curvature dipole are spatially modulated (see text). \textbf{b,} The net Berry curvature dipole along $k_y$ can be detected by a nonlinear Hall effect where the incident current is along $y$ and the resulting Hall voltage is along $x$.  \textbf{c,} Schematic measurement setup. \textbf{d,} Nonlinear transport measurement of UOTe at $T=250K$. \textbf{e,f,} Angular dependent nonlinear Hall measurements $T=250K$. The directions of both incident current $I^{^\omega}$ and the nonlinear Hall voltage $V^{2\omega}$ are rotated simultaneously with respect to the crystalline directions. The angular dependent measurements show large nonlinear Hall signal when the incident current is along $y$, revealing the directional Berry curvature dipole along $k_y$. \textbf{g,} The nonlinear Hall conductivity as a function of the Drude conductivity measured by temperature dependence between 200 K to 300 K.}
\label{Fig2}
\end{figure*}

\clearpage
\begin{figure*}[t]
\vspace{-0.5cm}
\includegraphics[width=14cm]{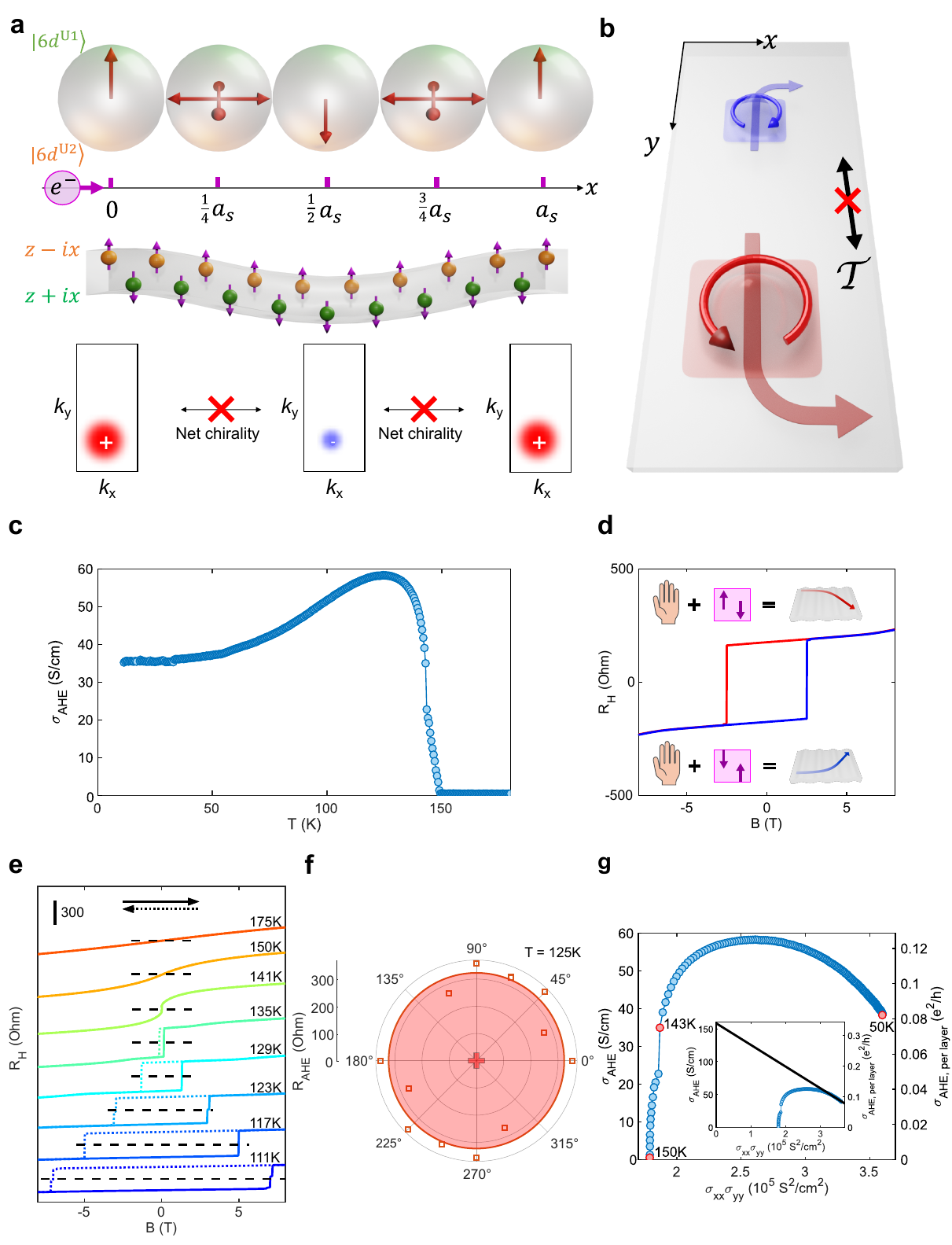}
\vspace{-0.5cm}
\caption{\textbf{Chiral superlattice plus AFM generates large anomalous Hall angle.} \textbf{a,} The collinear AFM order can break the cancellation between opposite Berry curvatures at $\pm k_y$: $x=0$ and $x=a_s/2$ have opposite total Berry curvature, which do not cancel due to the net chirality. \textbf{b,} The net Berry curvature in $k$ space generates the AHE. \textbf{c,} Temperature dependence of AHE conductivity. \textbf{d,} Magnetic field dependence of the Hall resistivity at $125$K, showing a pronounced anomalous Hall hysteresis loop. \textbf{e,} Temperature evolution of the anomalous Hall effect, showing a coercive field reaching up to $8$T near $110$K.  \textbf{f,} Angular dependence of the anomalous Hall resistance at $125$K.  \textbf{g,} Scaling behavior of AHE signal. The inset shows the fitting of $\sigma_{\text{AHE}} = \sigma_{\text{AHE}}^{\text{intrinsic}} + C^{\text{skew}}\sigma_{\text{Drude}}^2$ at $50\ \textrm{K}<T<100\ \textrm{K}$.}
\label{Fig3}
\end{figure*}

\clearpage
\begin{figure*}[t]
\vspace{-0.5cm}
\includegraphics[width=13cm]{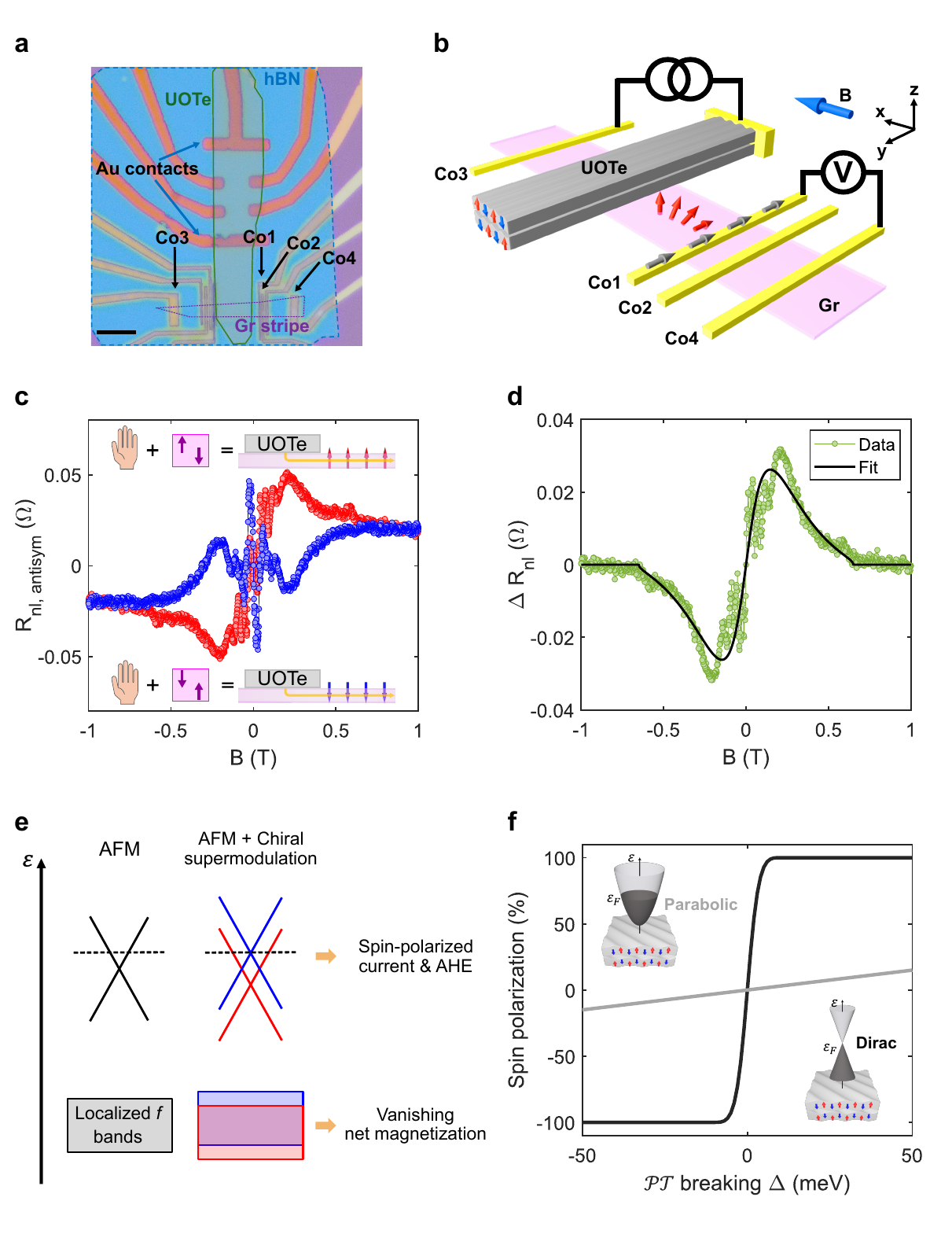}
\vspace{-0.5cm}
\caption{\textbf{Spin-polarized current from the chiral supermodulated collinear antiferromagnetism.} \textbf{a,b,} Schematic illustration and optical image of a nonlocal spin transport device consisting of a UOTe flake as the spin injector, Al$_2$O$_3$/Co contacts as the spin detectors, and a monolayer graphene channel for spin transport. The UOTe thickness is $27$ nm. The scale bar is $10$ $\mu$m. \textbf{c,} In-plane magnetic field dependence of the nonlocal resistance measured at 1.7K after initializing UOTe in AFM I and AFM II states while the Co detector magnetization held fixed. \textbf{d,} Hanle signal curve fitted to the solution of the time-independent Bloch equations for a uniform spin diffusion system (see details in main text). \textbf{e,} The small chiral imbalance means that the net magnetization of the U $5f$ will be small even if they cross the Fermi level. Moreover, in UOTe, the U $5f$ orbitals are filled so that they do not contribute to magnetization; The Fermi level electrons are small pockets whose contribution to the net magnetization is naturally small. On the other hand, the spin-polarized current is only determined by the spin polarization of the Fermi surface. UOTe features spin-split semimetal bands near Fermi level. The spin polarization of the current is naturally large. \textbf{f,} Model calculated spin polarization of the current as a function of the symmetry breaking $\Delta$.}
\label{Fig4}
\end{figure*}

\clearpage
\begin{figure*}[t]
\vspace{-0.5cm}
\includegraphics[width=15cm]{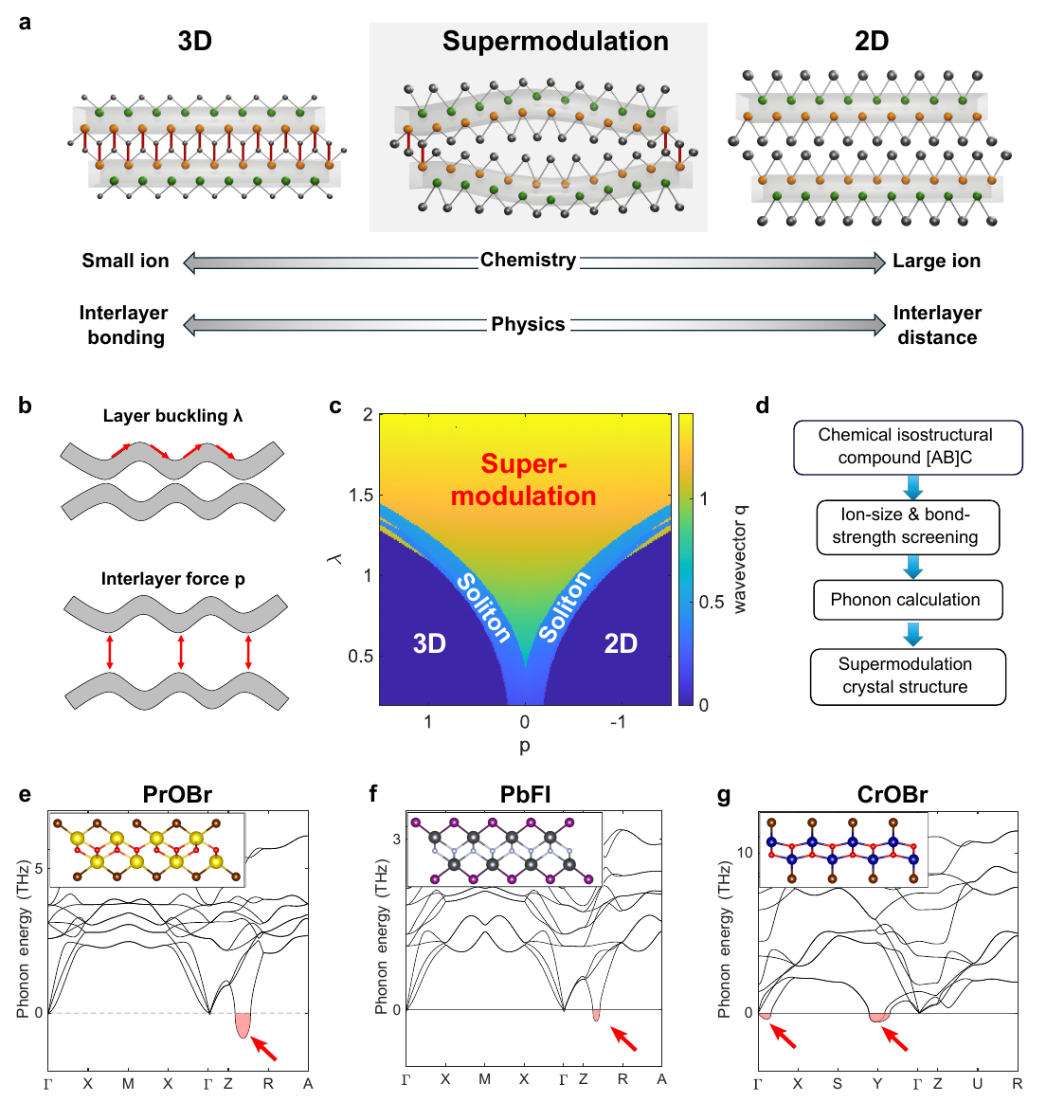}
\vspace{-0.5cm}
\caption{\textbf{Formation mechanism and design principle for chiral supermodulations.} \textbf{a,} The chiral supermodulation is an intermediate phase between the 3D covalent and 2D vdW phases due to competion between the interlayer bonding and intralayer repulsion energies that can be tuned by chemical ion sizes. \textbf{b,} Illustration of the layer buckling $\lambda$ and interlayer force $p$ that appear in the free energy. \textbf{c,}  Phase diagram as a function of  $\lambda$ and $p$, showing the supermodulation as an intermediate phase between 3D covalent and 2D vdW phases, with additional exotic soliton-superlattice phases at the boundaries between supermodulation and 2D/3D phases (see SI.II.1). \textbf{d,}  A flow chart for searching for similar chiral supermodulation in other materials. \textbf{e-g,}  DFT calculated phonon dispersions showing similar instability in other materials. }
\label{Fig5}
\end{figure*}

\setcounter{figure}{0}
\renewcommand{\figurename}{\textbf{Extended Data Fig.}}

\clearpage
\begin{figure*}[t]
\vspace{-0.5cm}
\includegraphics[width=18cm]{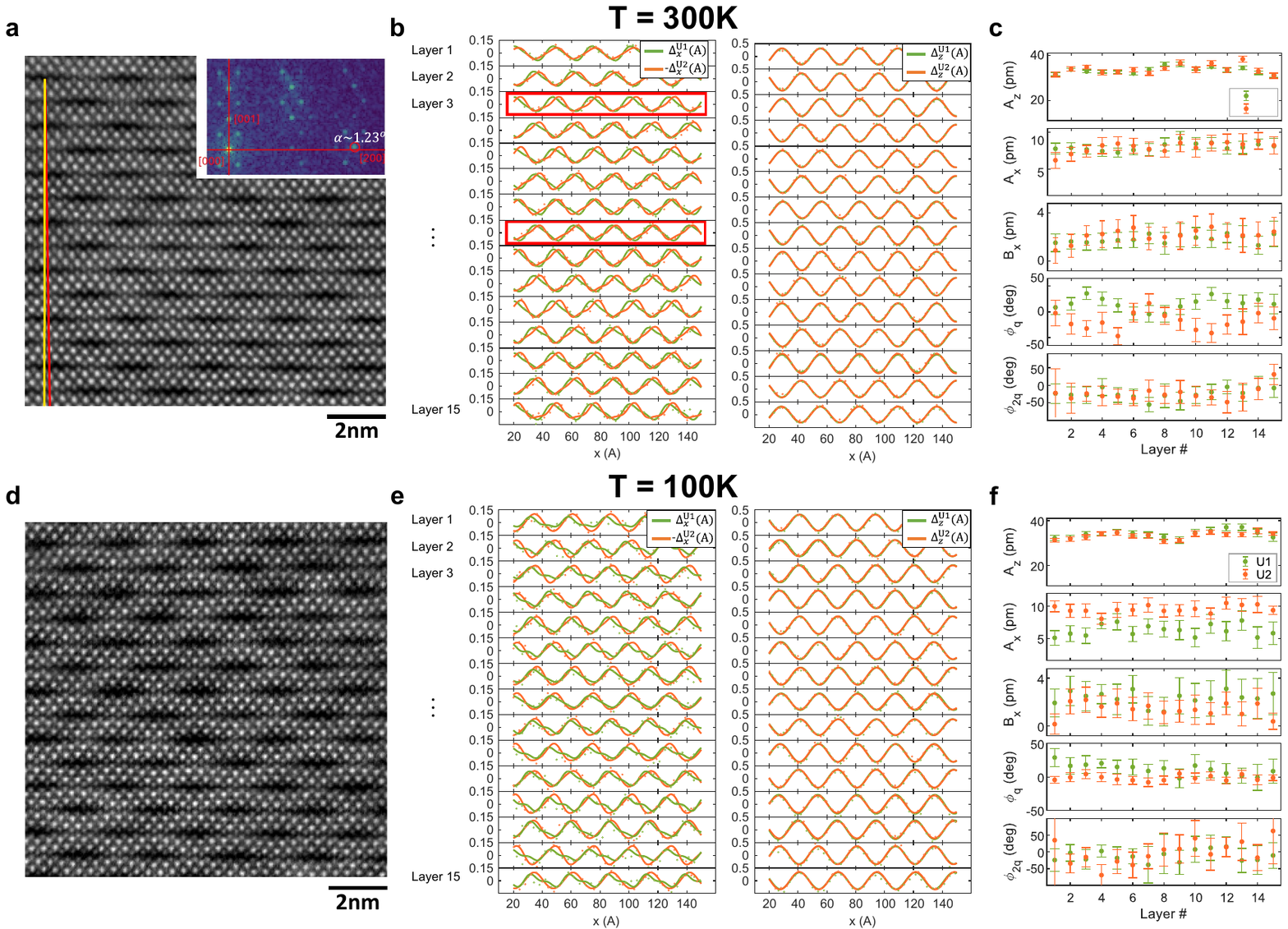}
\vspace{-0.0cm}
\caption{\textbf{Additional TEM at 300K and 100K in UOTe.} \textbf{a-c,} Additional TEM data at 300K. \textbf{a,} Cross-sectional STEM image of UOTe showing a monoclinic distortion. The red line connecting U atoms across layers deviates from the vertical yellow reference line, visualizing the cumulative shear associated with the monoclinic distortion. Inset: Electron diffraction pattern acquired from the same cross-sectional region, revealing a monoclinic distortion angle of $\alpha \sim 1.23^o$. \textbf{b,} Fitted displacement parameters of U atoms across 15 UOTe layers. See Methods.3 for the expressions for fitting the TEM data. The in-plane displacements of different layers have different lineshapes (for example, one can compare the highlighted red boxes). Therefore $t\mathcal{T}$ symmetry is also broken. \textbf{c,} Fitted values for each layer. \textbf{d-f,} Same as panels \textbf{a-c} but for TEM data at 100K.}
\label{Extended_TEM_300K_100K}
\end{figure*}

\clearpage
\begin{figure*}[t]
\vspace{-0.5cm}
\includegraphics[width=17.5cm]{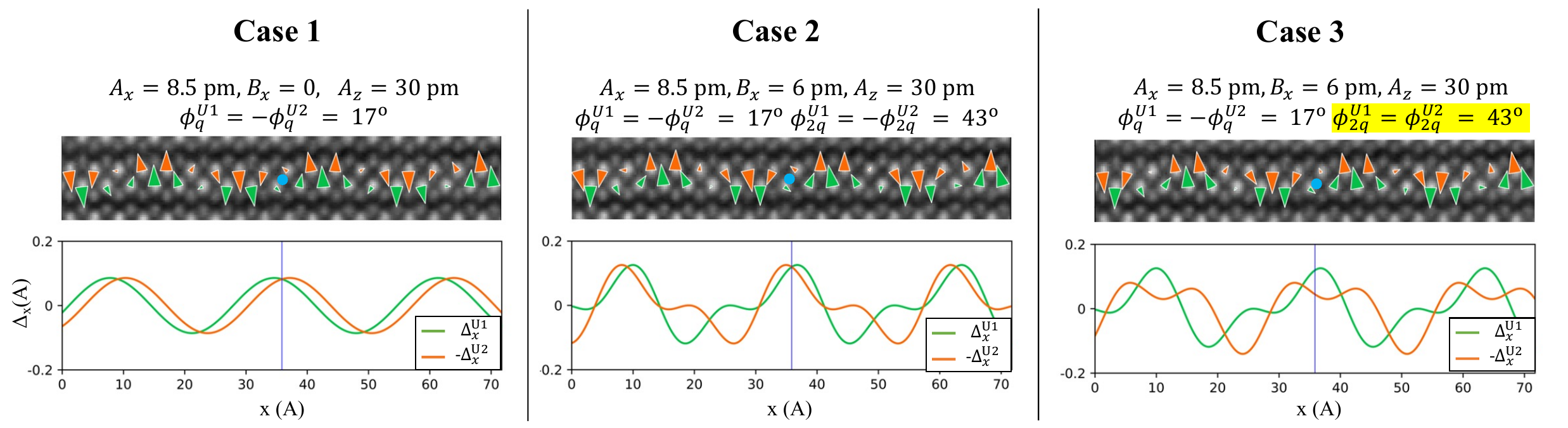}
\vspace{-0.0cm}
\caption{\textbf{Simulations of the TEM displacement under different amplitudes and phases.} Case 1: If we only turn on the first harmonic $z$ and first harmonic $x$ modulations with opposite phases. Indeed, the $x+iz$ (orange) and $x-iz$ (green) modulations are perfectly symmetric with respect to the inversion center (the blue circle). The in-plane displacement also shows two simple sine waves that are symmetric with respect to each other. Case 2: if we now add second-harmonic $x$ modulation with opposite phases. Inversion is preserved except that the line-shape is more complex. Case 3: If we now change the second harmonic modulation of U1 and U2 to have the same phase, then inversion is broken. }
\label{Extended_Structural_Simulations}
\end{figure*}

\clearpage
\begin{figure*}[t]
\vspace{-0.5cm}
\includegraphics[width=11cm]{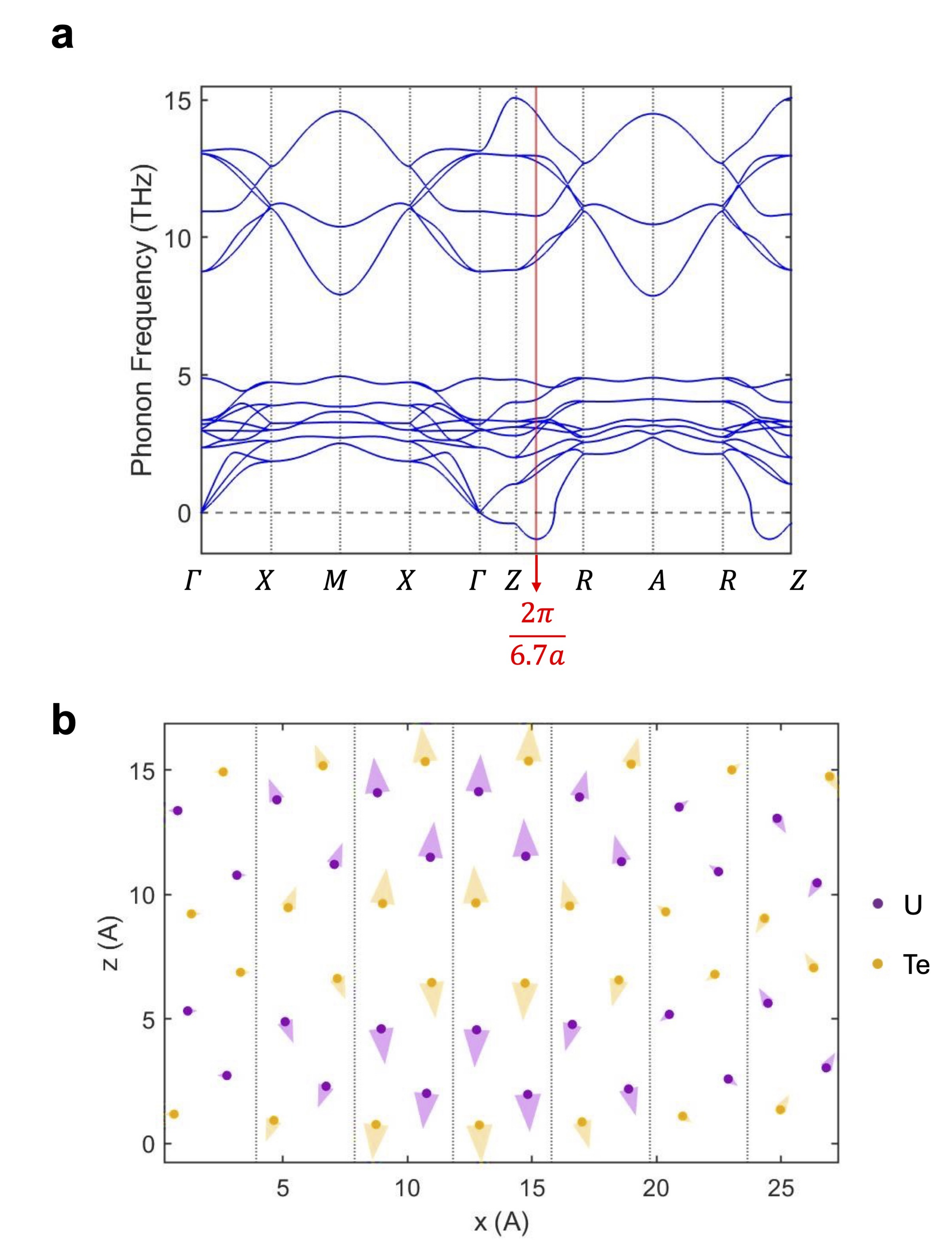}
\vspace{-0.0cm}
\caption{\textbf{DFT calculated phonon band structure and displacement of UOTe without superlattice.} \textbf{a,} DFT calculated phonon band structure of UOTe. One branch of optical phonon has negative frequency at $\vec{q}=(\frac{1}{6.7}\frac{2\pi}{a}, 0, \frac{1}{2}\frac{2\pi}{c}$), consistent with the experimentally observed $\vec{q}$ for the superlattice.. \textbf{b,} Atomic displacements corresponding to that unstable phonon at the negative frequency. U1 and U2 (also Te1 and Te2) have opposite chiral winding structures, consistent with the experimental findings.}
\label{Extended_DFT_phonon}
\end{figure*}

\clearpage
\begin{figure*}[t]
\vspace{-0.5cm}
\includegraphics[width=17.5cm]{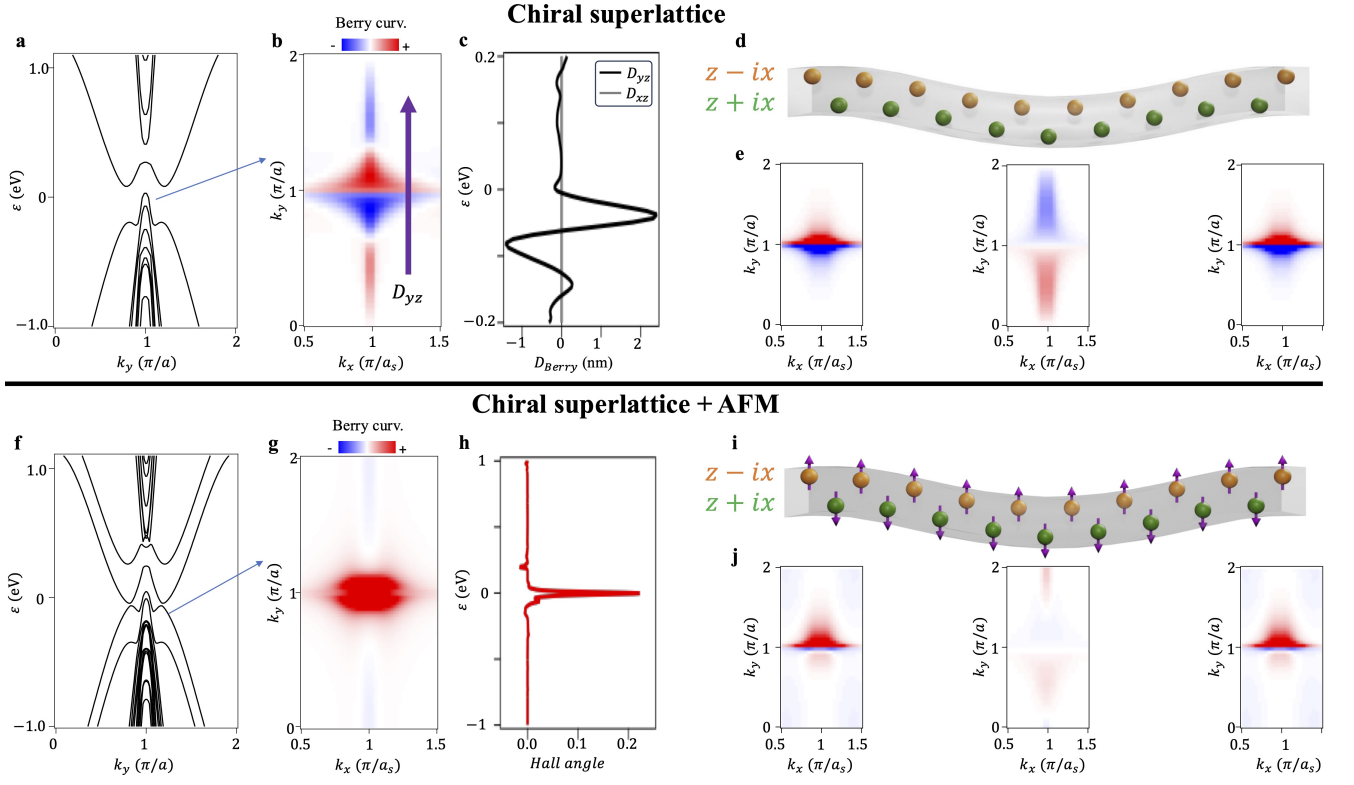}
\vspace{-0.0cm}
\caption{\textbf{Theoretical tight-binding modeling of the chiral superlattice quantum geometry.} \textbf{a,} Chiral superlattice band structure. \textbf{b,} Berry curvature $k$-space distribution of the low-energy superlattice band. \textbf{c,} Calculated Berry curvature dipole. \textbf{d,} Schematic of the chiral superlattice containing the $z+ix$ U1 and $z-ix$ U2 chiral displacements. \textbf{e,} Spatially-resolved Berry curvature at different $x$ values in the superlattice, calculated by treating the slowly varying superlattice modulation as an adiabatic parameter and approximating each local region by a homogeneous tight-binding Hamiltonian $H(\mathbf{k};x)$. \textbf{f-j,} The same as panels (\textbf{a-e}) but for chiral superlattice plus the AFM order. Panel \textbf{h} shows the calculated AHE Hall angle.}
\label{Extended_Berry_curv_THY_AFM}
\end{figure*}

\clearpage
\begin{figure*}[t]
\vspace{-0.5cm}
\includegraphics[width=17.5cm]{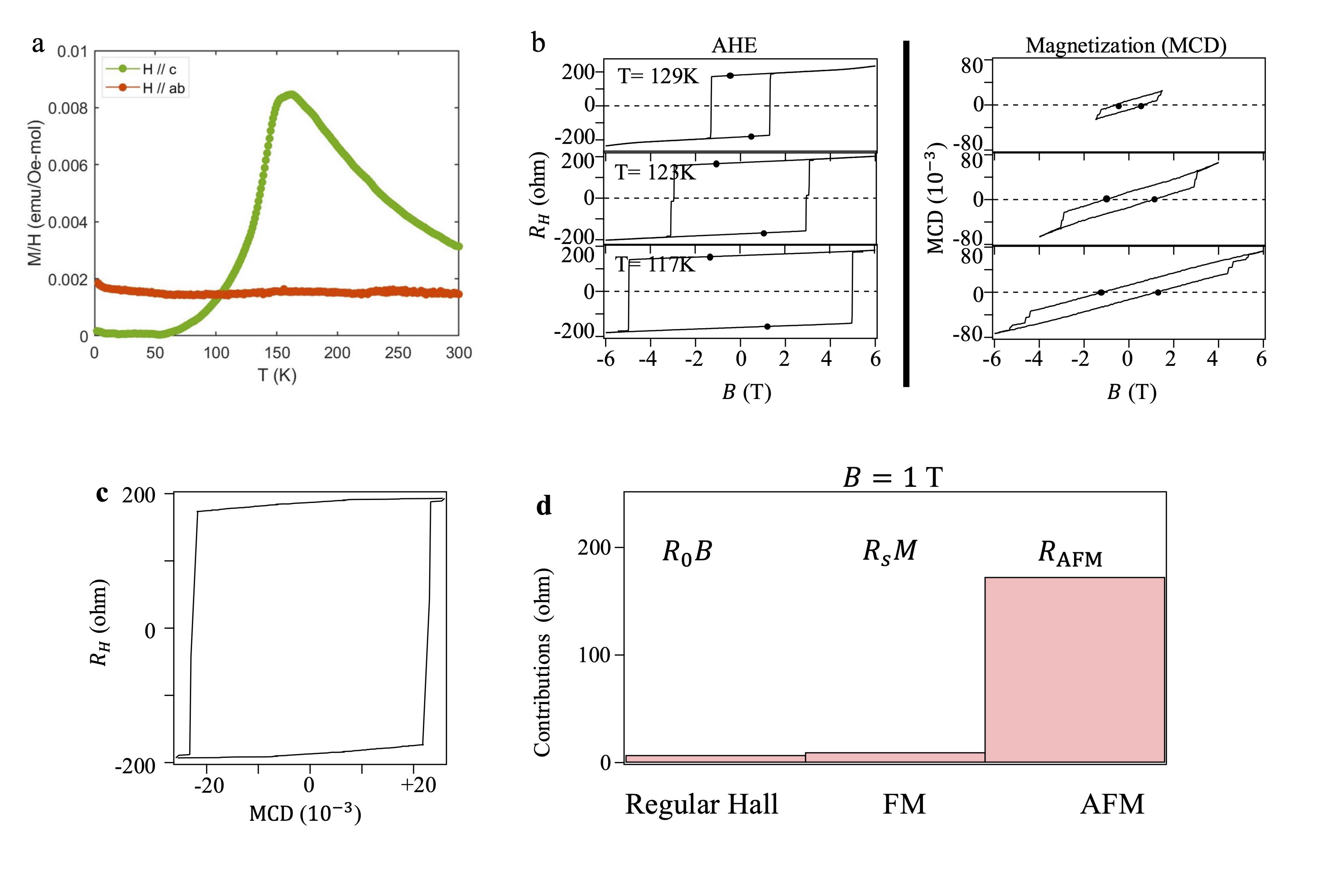}
\vspace{-0.0cm}
\caption{\textbf{Excluding weak ferromagnetism and regular Hall effect as the main contribution of the AHE.} \textbf{a,} Temperature dependence of in-plane and out-of-plane magnetic susceptibilities ($\chi_{//}$ and $\chi_{\perp}$). \textbf{b,} AHE and magnetization (measured by magnetic circular dichroism). The magnetic circular dichroism is the antisymmetric part of the optical circular dichroism, which naturally excludes the chiral circular dichorism (see details in SI.III.3). The net magnetization under finite magnetic field is given by $M=M_\textrm{s}+\chi_{\perp}B$.$M_\textrm{s}$ is the spontaneous magnetization which was measured by the NV center magnetometry to be $\sim10$ m$\mu_{\textrm{B}}$ per U ion. Due to the small $M_\textrm{s}$ and large coercive field, large AHE is observed at zero magnetization, therefore demonstrating that the small net magnetization is a side product of the symmetry breaking. \textbf{c,} AHE as a function of the magnetization using the data in panel (\textbf{b}) for 129 K. \textbf{d,}  We can further discern the different contributions to the Hall effect including the regular Hall effect, the weak ferromagnetism and the intrinsic AFM, i.e., $R_\textrm{AHE}=R_0B+ R_\textrm{s}M_\textrm{s}+R_\textrm{AFM}$. $R_0$ and $R_s$ are slopes of the AHE vs. B field and AHE vs. magnetization curves. The different contributions at 129 K is shown here.  The intrinsic AFM contribution strongly dominates.}
\label{Extended_Excluding_M}
\end{figure*}

\clearpage
\begin{figure*}[t]
\vspace{-0.5cm}
\includegraphics[width=14cm]{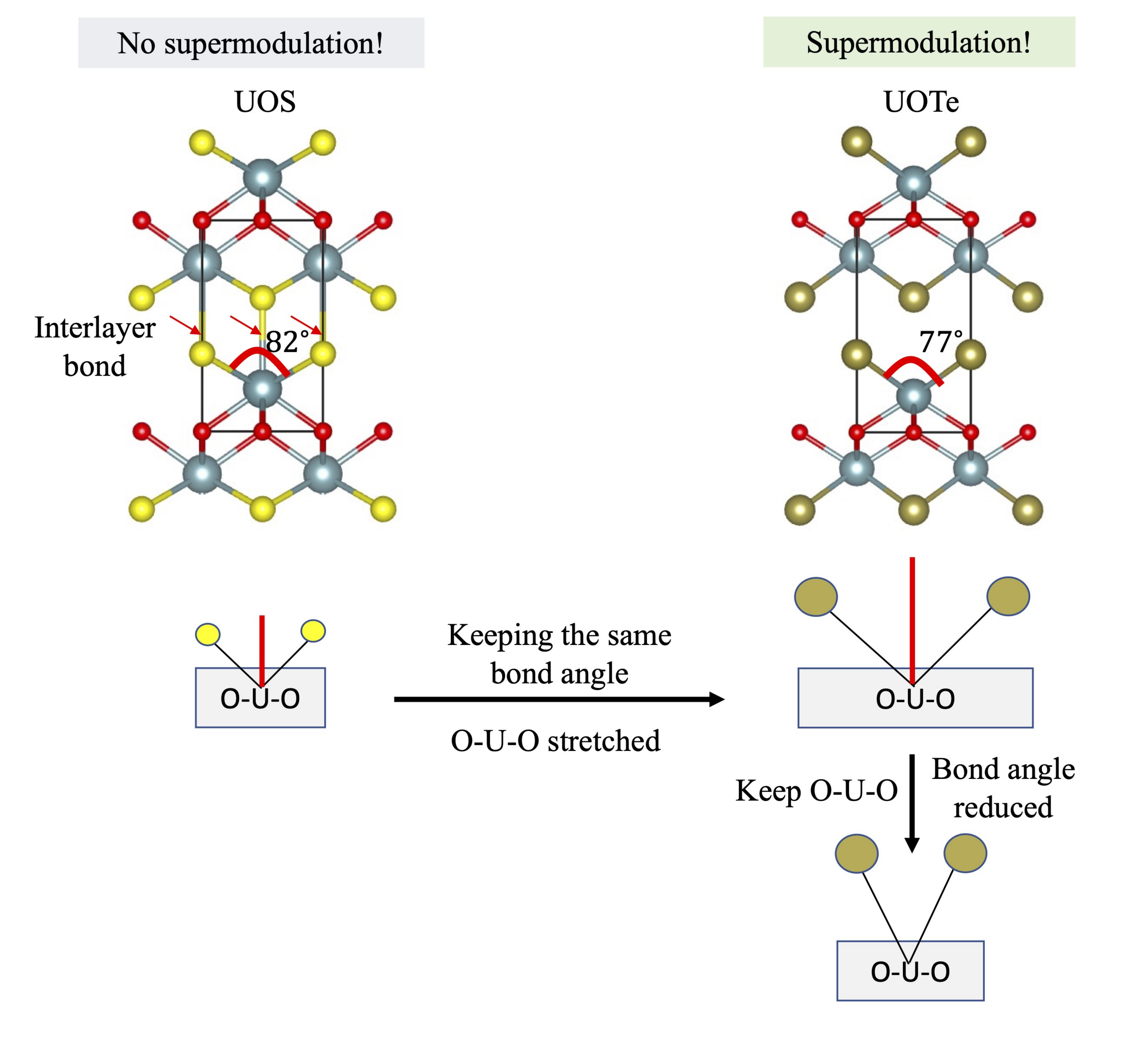}
\vspace{-0.0cm}
\caption{\textbf{Superlattice formation mechanism 1 - A comparison between UOS and UOTe.} These two compounds are isostructural, but UOS has no supermodulation. A comparison of the nominal crystal structure shows that  UOS has an additional interlayer bond connecting the U with the S in the next layer. This is enabled by the large S-U-S angle ($82$ deg), which resembles opening arms. Such an interlayer bond lowers energy. Forming similar interlayer bonding in UOTe is too energy costly as the much longer  U-Te bond length would laterally stretch the U-O-U central at similarly large bond angle. As a result, in the nominal unit cell of UOTe, the Te-U-Te bond angle is much smaller ($77$ deg), protecting the U-O-U central block from being stretched but in turn hindering the interlayer bonding.}
\label{Superlattice_Mechanism_1}
\end{figure*}

\clearpage
\begin{figure*}[t]
\vspace{-0.5cm}
\includegraphics[width=13.5cm]{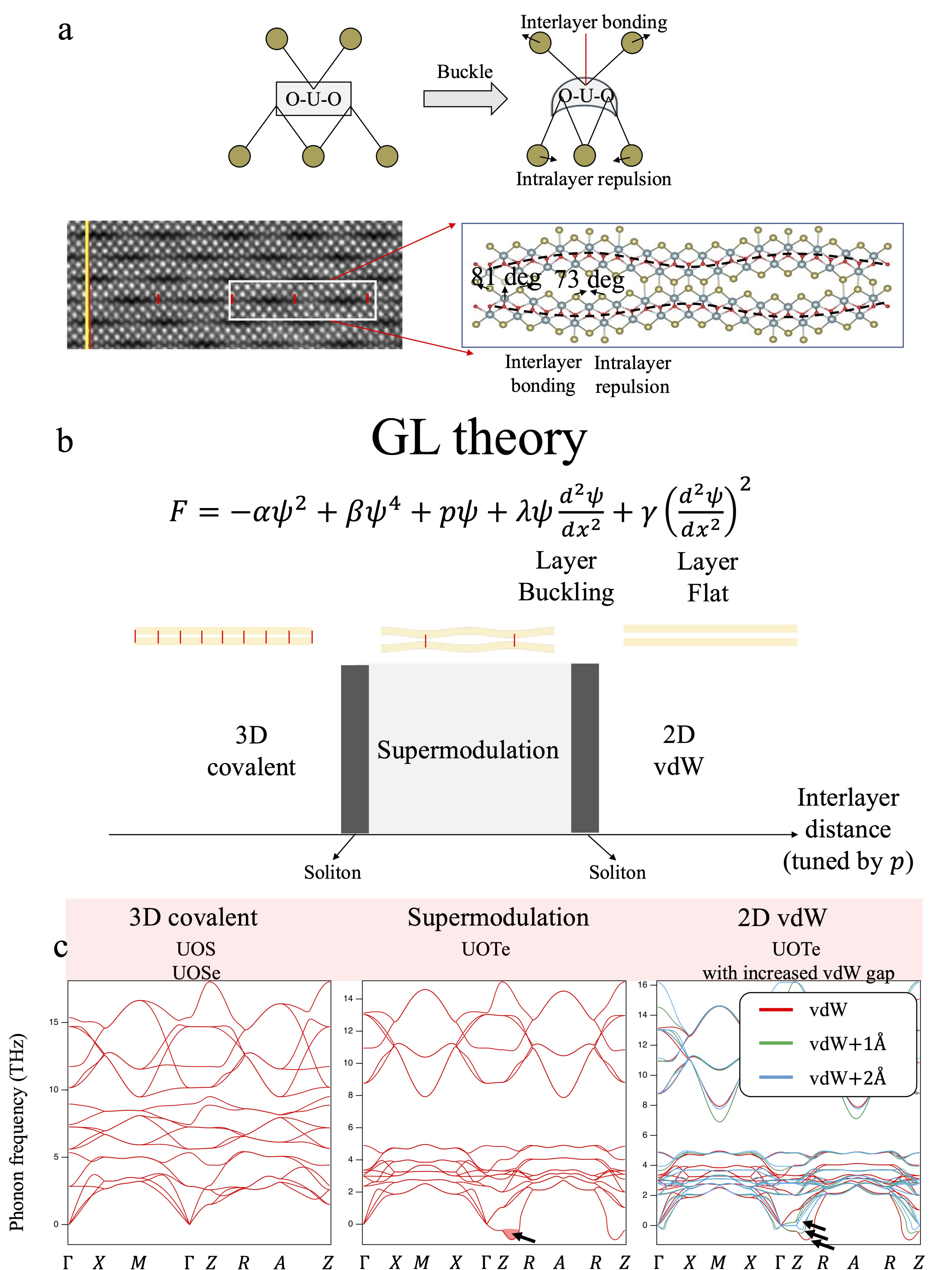}
\vspace{-0.0cm}
\caption{\textbf{Superlattice formation mechanism 2 - Supermodulation as an intermediate phase.} \textbf{a,} To form the interlayer bonding while keeping the central block unstretched, the layers decide to locally curve up/down with opposite sign in adjacent layers, which is the supermodulation. In this way, at specific locations, the Te-U-Te angle opens up ($81$ deg), allowing the interlayer bonding. At some other locations, the Te-U-Te angle reduces ($73$ deg), leading to an intralayer repulsion. The competition between the interlayer bonding and intralayer repulsion gives rise to the supermodulation in UOTe. \textbf{b,} More generally, we can consider three distinct phases: (1) 3D covalent, (2) 2D vdW, (3) supermodulation as an intermediate phase. The phase diagram is described by a Ginzburg Landau theory (see main text). \textbf{c,} DFT calculated phonon dispersions of UOS, UOTe and UOTe with artificially increased vdW gap.}
\label{Superlattice_Mechanism_2}
\end{figure*}

\clearpage
\begin{figure*}[t]
\vspace{-0.5cm}
\includegraphics[width=14.5cm]{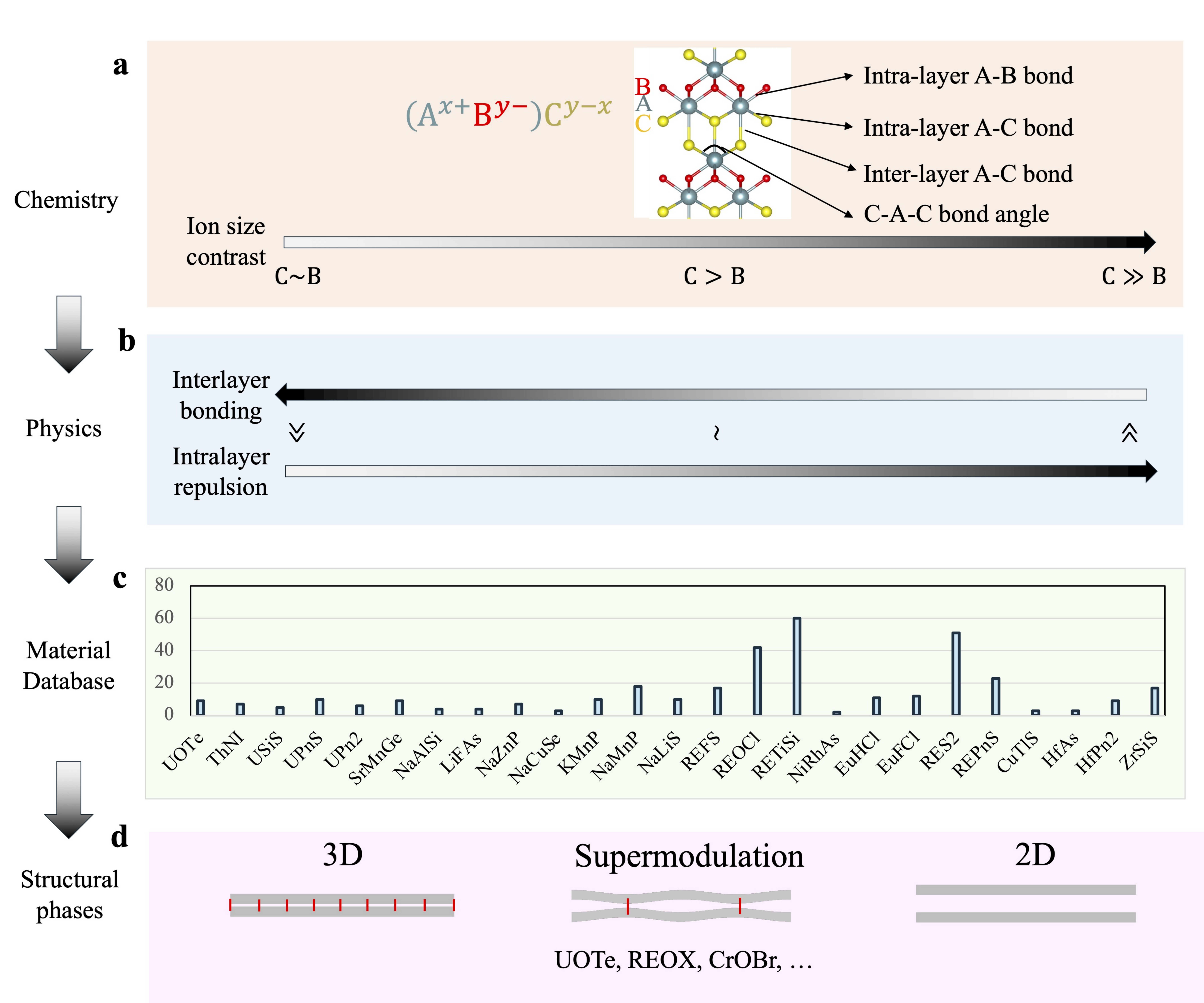}
\vspace{-0.0cm}
\caption{\textbf{Proposed design principle for chiral supermodulation.} Based on the above mechanism and phenomenological theory, we propose a design principle that goes from chemistry to physics to materials. \textbf{a,} From chemistry, we can tune the size of the ions and the strength of chemical bonds.  \textbf{b,} To physics, the above chemical tuning of ion size and bond strength effectively controls the interlayer bonding and intralayer repulsion energies.  \textbf{c,} To materials, there are $\sim500$ compounds isostructural to UOTe in the ICSD database. We have developed an empirical rule based on material parameters such as the ion size, the C-A-C bond angle, and the ratios of bond lengths including $r_1=\frac{\textrm{Inter-layer A-C bond}}{\textrm{Intra-layer A-C bond}}$ and $r_2=\frac{\textrm{Intra-layer A-C bond}}{\textrm{Intra-layer A-B bond}}$. \textbf{d,} Based on empirical rules, we can sift through the material database to identify the most promising materials (see SI.II.4). }
\label{Design_principle}
\end{figure*}

\clearpage
\begin{figure*}[t]
\vspace{-0.5cm}
\includegraphics[width=14cm]{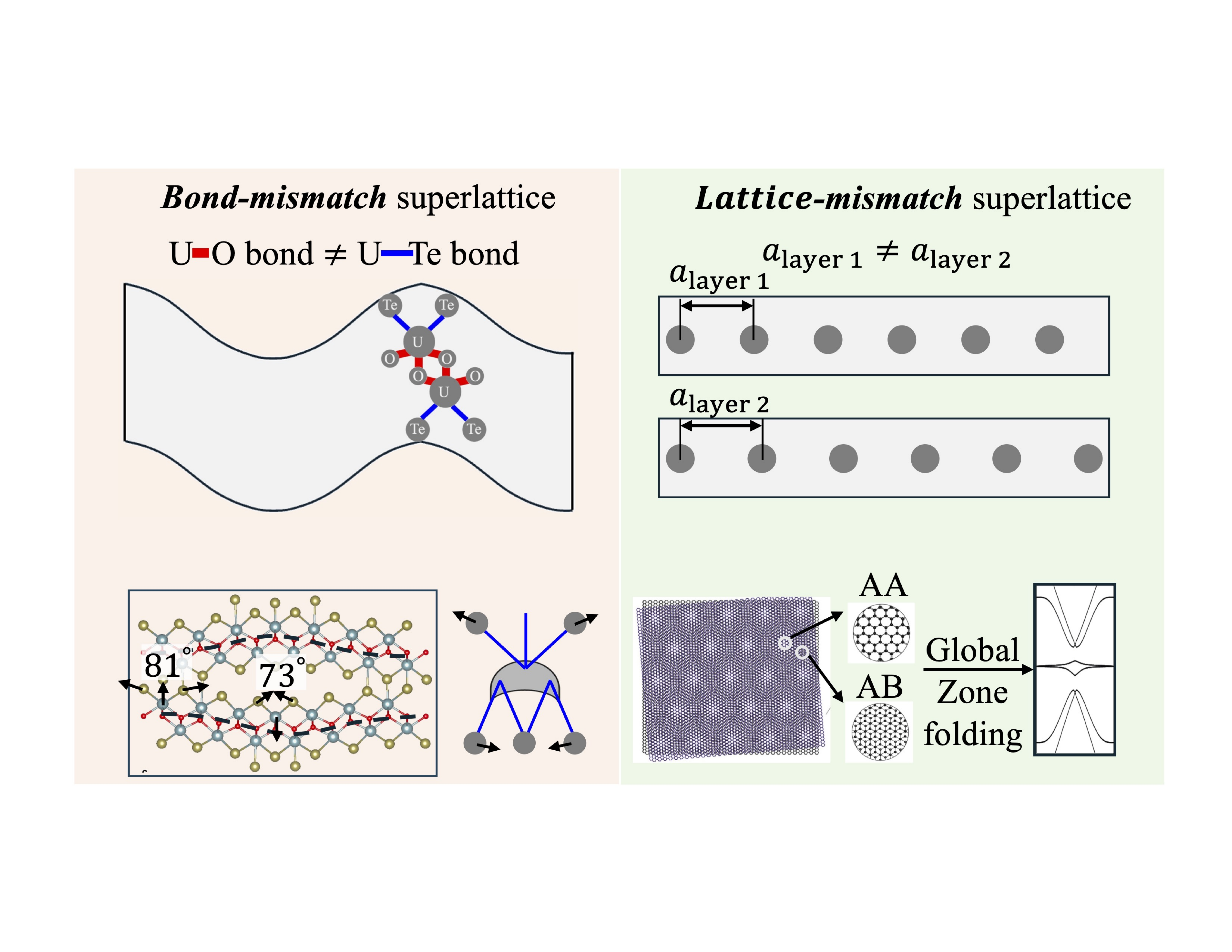}
\vspace{-0.0cm}
\caption{\textbf{Bond-mismatch superlattice and lattice-mismatch superlattice.} UOTe is a bond-mismatch superlattice, in which the superlattice strongly modifies the local U-O and U-Te bonding environment itself. This provides a real-space handle on the orbital composition and hopping amplitudes, which in turn produces spatially modulated Berry curvature, Berry-curvature dipole, and related quantum-geometric responses (see SI.II.5). }
\label{Extended_Superlattice_concept}
\end{figure*}

\end{document}